\documentclass[sn_mathphys_num]{sn_jnl}

\usepackage{lmodern}

\usepackage{graphicx}%
\usepackage{multirow}%
\usepackage{amsmath,amssymb,amsfonts}%
\usepackage{amsthm}%
\usepackage{mathrsfs}%
\usepackage[title]{appendix}%
\usepackage{xcolor}%
\usepackage{textcomp}%
\usepackage{manyfoot}%
\usepackage{booktabs}%
\usepackage{algorithm}%
\usepackage{algorithmicx}%
\usepackage{algpseudocode}%
\usepackage{listings}%
\usepackage[numbers]{natbib}
\usepackage[justification=centering]{caption} 
\usepackage{subcaption}
\usepackage{placeins}

\begin{document}

\title[Article Title]{Advances in  Modeling Techniques for Ventricular Assist Devices: A Comprehensive Review and Future Directions}

\author{
Muhammad Adel Yusuf\textsuperscript{1},
Nezar M. Alyazidi\textsuperscript{1,2},
Hamna Saleem\textsuperscript{6},
Ali Nasir\textsuperscript{1,3,5,*},
Mojeed Oyedeji\textsuperscript{4}
}
\maketitle
\begin{center}
\textsuperscript{1}Department of Control and Instrumentation Engineering, King Fahd University of Petroleum and Minerals, Dhahran 31261, Saudi Arabia. \\
\textsuperscript{2}Interdisciplinary Research Center for Smart Mobility and Logistics, King Fahd University of Petroleum and Minerals, Dhahran 31261, Saudi Arabia. \\
\textsuperscript{3}Interdisciplinary Research Center for Intelligent Manufacturing and Robotics (IRC-IMR), King Fahd University of Petroleum and Minerals, Dhahran 31261, Saudi Arabia. \\
\textsuperscript{4}SDAIA-KFUPM Joint Research Center for Artificial Intelligence, King Fahd University of Petroleum and Minerals, Dhahran 31261, Saudi Arabia.\\
\textsuperscript{5}Interdisciplinary Research Center for Aviation and Space Exploration (IRC-ASE), King Fahd University of Petroleum and Minerals, Dhahran 31261, Saudi Arabia.\\
\textsuperscript{6}University of Engineering \& Technology, Lahore, Pakistan.\\
*Correspondence to: ali.nasir@kfupm.edu.sa\\

\end{center}

\begin{abstract}
    
\vspace{1em}

\begin{center}
\textbf{Abstract}
\end{center}

\vspace{0.5em}

Ventricular Assist Devices (VADs), particularly rotary Left Ventricular Assist Devices (LVADs), have become an essential therapy for patients with advanced heart failure who are ineligible for heart transplantation. Despite advances in cardiovascular modeling and control strategies, the clinical translation of proposed LVAD control approaches remains limited. This review critically examines the evolution of LVAD modeling and control methods and argues that this gap arises not from a lack of sophisticated algorithms, but from mismatches between modeling assumptions, sensing constraints, and real-world cardiovascular variability.

A systematized review of over 100 peer-reviewed studies is conducted, covering mathematical models of the cardiovascular system, lumped-parameter and reduced-order representations, classical and advanced control strategies, and emerging data-driven and machine learning based approaches. The reviewed literature is synthesized through a problem-driven framework that links modeling and control choices to clinical challenges, including physiological observability, parameter identifiability, robustness to patient variability, and prevention of adverse events such as ventricular suction and thrombosis.

The analysis reveals that while high-fidelity models and intelligent control strategies demonstrate performance in simulation, their reliance on unmeasurable states, extensive parameter tuning, and dense sensing limits implementation. Simpler control approaches often exhibit greater robustness under clinical constraints. Emerging adaptive and data-driven techniques show potential to bridge this divide, but only when designed with implantable sensing limitations and interpretability requirements in mind.

By identifying the structural barriers that hinder adoption, this review provides a synthesis of LVAD modeling and control paradigms and outlines research directions aimed at achieving physiologically adaptive, clinically implementable, and patient-safe LVAD systems.
\end{abstract}


\section{Introduction}
Heart failure (HF) affects more than 2\% of the global population and remains a leading cause of morbidity and mortality worldwide \cite{yin2024real}. In the United States alone, over 6.5 million individuals live with HF, resulting in approximately 700,000 deaths annually \cite{tsao2022heart}. Although heart transplantation represents the definitive treatment for end-stage HF, only a small fraction of patients are eligible due to donor scarcity and medical contraindications \cite{yin2024real}. Consequently, mechanical circulatory support devices—particularly Left Ventricular Assist Devices (LVADs)—have become a critical therapeutic option for improving survival and quality of life in patients with advanced HF.

Modern LVADs mechanically unload the failing left ventricle by drawing blood from the ventricular apex and delivering it to the aorta, functioning either as a bridge to transplantation or as destination therapy \cite{slaughter2009advanced}. While early pulsatile LVADs were designed to mimic physiological cardiac rhythms, their large size and mechanical wear limited long-term reliability \cite{raman2004destination}. Continuous-flow rotary LVADs, based on axial or centrifugal pump designs, have since become the clinical standard due to their compact size, durability, and improved energy efficiency \cite{caccamo2011current, carpenter2013brief, loor2012pulsatile}. Despite these advances, contemporary LVADs lack intrinsic physiological regulation and cannot autonomously adapt to changes in preload, afterload, or patient activity, making effective modeling and control essential for safe long-term support \cite{son2020modelling}.

Over the past two decades, extensive research efforts have focused on developing mathematical models and control strategies to regulate LVAD operation. Cardiovascular system (CVS) representations ranging from time-varying elastance and Windkessel models to lumped-parameter and higher-order reduced-order models have been proposed to capture LVAD-CVS interactions. In parallel, control approaches spanning classical methods such as Proportional-Integral-Derivative (PID) control to more advanced techniques including Sliding Mode Control (SMC), adaptive control, and artificial intelligence (AI)-based strategies have been investigated. However, despite the increasing sophistication of these methods, relatively few have achieved widespread clinical adoption, and many remain confined to simulation or benchtop validation.

This review argues that the limited clinical translation of advanced LVAD modeling and control strategies does not primarily stem from insufficient algorithmic capability, but rather from fundamental structural constraints. Specifically, three interconnected bottlenecks persist: (i) limited physiological observability due to the absence of reliable implantable sensors, (ii) parameter identifiability challenges arising from patient-specific and time-varying cardiovascular dynamics, and (iii) a mismatch between model or controller complexity and clinical implementability. These constraints contribute directly to adverse clinical outcomes such as ventricular suction, thrombosis, and inadequate cardiac unloading, highlighting the need for a more critical examination of how modeling and control assumptions influence real-world performance.

Existing review articles have examined cardiovascular modeling techniques \cite{shim2004mathematical, kokalari2013review, capoccia2015development, vseman2024computational}, LVAD control strategies \cite{alomari2012developments}, optimization methods \cite{suero2023optimization}, and AI-based applications \cite{balcioglu2024role, grzyb2024artificial}. However, most reviews remain method-centric, providing descriptive summaries of models or algorithms without critically analyzing their translational limitations or interdependencies. As a result, key trade-offs such as model fidelity versus parameter identifiability, or control optimality versus robustness under sensing constraints are often underexplored, limiting the utility of these reviews for interdisciplinary research and clinical decision-making.

To address these gaps, this review presents a problem-driven synthesis of LVAD modeling and control literature through a translational framework that links modeling choices and control strategies to clinical constraints and failure modes. Particular emphasis is placed on lumped-parameter modeling approaches, control robustness under limited observability, and the emerging role of data-driven and machine learning-based methods when designed within realistic sensing and interpretability boundaries. By shifting the focus from cataloging techniques to analyzing why certain approaches succeed or fail in practice, this review aims to provide actionable insights for engineers, researchers, and clinicians working toward physiologically adaptive, clinically implementable, and patient-safe LVAD systems.

The remainder of this paper is organized as follows. Section~\ref{Modeling} examines cardiovascular and LVAD modeling techniques through the lens of fidelity, identifiability, and clinical feasibility. Section~\ref{Control} analyzes control strategies with emphasis on robustness, sensing requirements, and failure prevention. Section~\ref{AI} discusses simulation-based and data-driven approaches, highlighting opportunities and limitations of machine learning under implantable constraints. Section~\ref{Future} outlines research directions grounded in unresolved translational challenges, and Section~\ref{Conclusion} concludes the paper.

\section{Modeling}
\label{Modeling}
\label{Mathematical modeling}

\begin{figure*}[t!]
\centering
\includegraphics[width=0.8\textwidth]{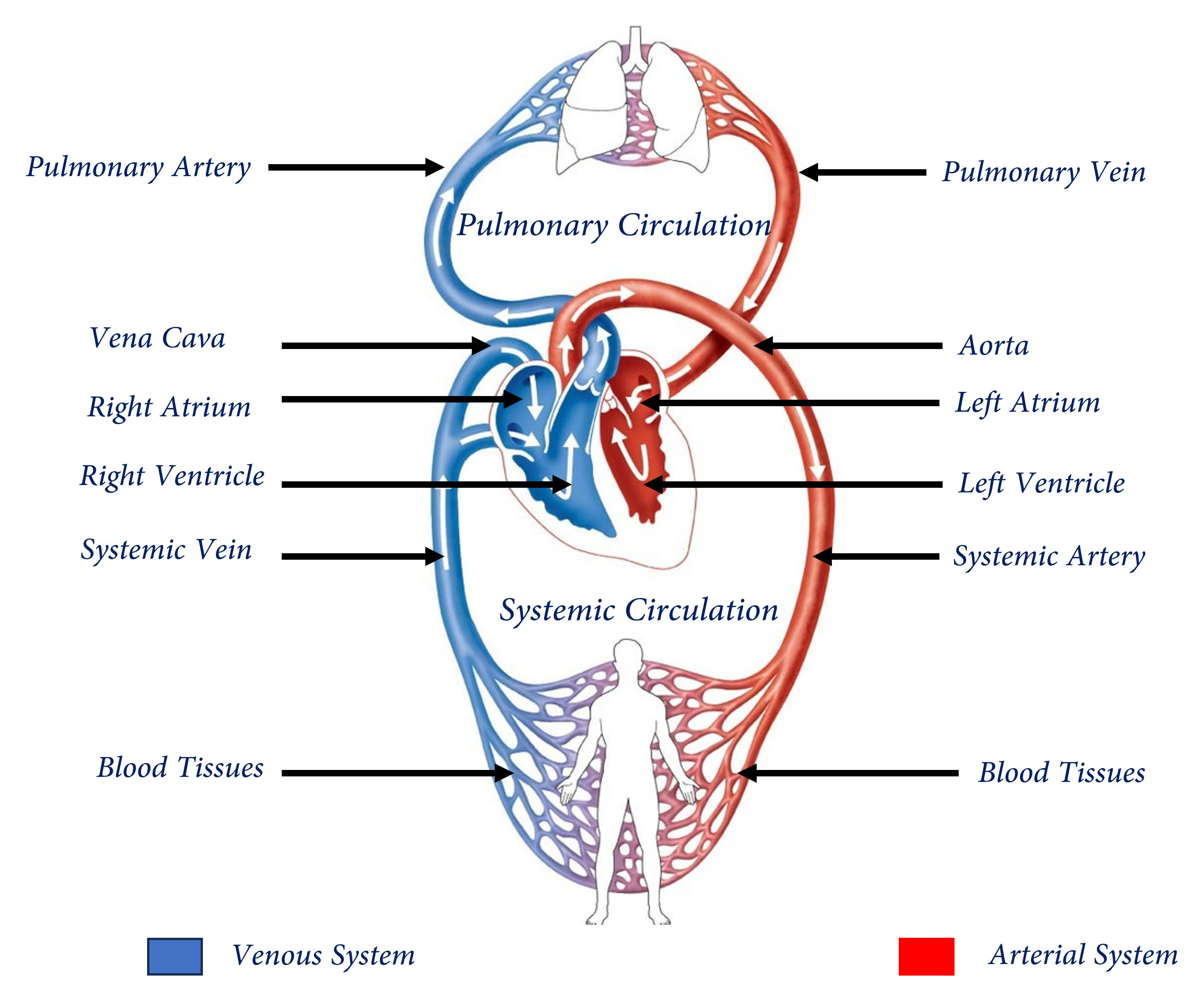}
\caption{Schematic illustration of the CVS}\label{CVS}
\end{figure*}
The cardiovascular system (CVS) is a highly nonlinear, closed-loop dynamical system driven by the heart and regulated through complex mechanical, biochemical, and neural feedback mechanisms. Its primary function is to transport oxygen, nutrients, metabolic waste, and heat via blood circulation while maintaining tightly regulated pressure and flow conditions across a wide range of physiological states. Structurally, the CVS consists of the heart, systemic circulation, pulmonary circulation, and regulatory subsystems, whose coordinated interaction ensures circulatory homeostasis. A schematic overview of these components is shown in Figure~\ref{CVS}.

At its core, the heart operates as a dual pump, with the right atrium and ventricle propelling deoxygenated blood to the pulmonary circulation and the left atrium and ventricle delivering oxygenated blood to the systemic circulation. Unidirectional flow is enforced by four cardiac valves: tricuspid, pulmonary, mitral, and aortic, whose dynamics introduce additional nonlinearities into the system. Systemic and pulmonary vasculatures further contribute resistive, compliant, and inertial effects, while biochemical and autonomic regulatory mechanisms dynamically adjust heart rate, vascular resistance, and blood volume in response to metabolic demand and pathological conditions.

From a physical perspective, blood flow within the CVS obeys the conservation laws of mass, momentum, and energy. However, vessel compliance, time-varying boundary conditions, and strong coupling between pressure and flow significantly complicate analytical treatment. These challenges are amplified when mechanical circulatory support devices such as left ventricular assist devices (LVADs) are introduced. Unlike the native heart, LVADs generate continuous or semi-continuous flow and lack intrinsic physiological feedback, fundamentally altering ventricular loading conditions, pulsatility, and pressure-volume relationships. As a result, accurate modeling of LVAD–CVS interaction is not merely descriptive but essential for predicting adverse events such as suction, ventricular collapse, thrombosis, and impaired end-organ perfusion.

Mathematical modeling has therefore become an indispensable tool for studying CVS hemodynamics, enabling quantitative analysis of pressure–flow distributions, evaluation of device–heart interactions, and in silico testing of control strategies prior to clinical deployment \cite{fu2015influence, formaggia2010cardiovascular}. Importantly, models serve different purposes depending on their level of abstraction: high-fidelity models aim to reproduce physiological detail, whereas reduced-order models prioritize computational efficiency, interpretability, and suitability for real-time control. This trade-off between physiological realism and clinical practicality lies at the heart of contemporary LVAD modeling research and remains a central unresolved challenge.

Existing cardiovascular and LVAD models can be broadly classified into two categories: deterministic and stochastic. Deterministic models assume known system dynamics and parameters, providing clear mechanistic insight into ventricular mechanics, vascular behavior, and pump-heart coupling. These models form the backbone of most LVAD simulations and control designs but often struggle to accommodate patient-specific variability, measurement uncertainty, and long-term physiological adaptation. In contrast, stochastic models explicitly account for uncertainty and variability in system parameters, measurements, or external disturbances, offering a probabilistic framework that is better aligned with real-world clinical conditions but typically at the cost of increased complexity and reduced interpretability.

In the context of LVAD research, the choice of modeling paradigm has direct implications for control robustness, sensor requirements, and clinical translatability. Deterministic models enable transparent analysis and controller design but rely heavily on parameters that are difficult or impossible to measure in vivo. Stochastic and data-driven extensions promise improved adaptability yet raise questions regarding validation, safety, and regulatory acceptance. Understanding the strengths and limitations of these modeling approaches, and how they can be combined, is therefore essential for advancing LVAD technology beyond laboratory prototypes toward reliable clinical systems.

The following subsections critically examine deterministic and stochastic modeling approaches used in LVAD-supported CVS studies. Rather than presenting a catalog of models, the focus is placed on their underlying assumptions, practical limitations, and relevance to real-time control and clinical deployment. This analysis provides the foundation for later discussions on intelligent control strategies and future research directions.

\subsection{Deterministic Models}
Within the deterministic modeling paradigm introduced in the previous section, a wide range of model structures have been proposed to study LVAD-supported cardiovascular dynamics. These models differ primarily in how they represent ventricular mechanics, vascular behavior, and pump-heart coupling, reflecting distinct priorities in terms of physiological detail, computational efficiency, and suitability for control design.

In LVAD research, deterministic models are most commonly employed to analyze ventricular unloading, assess hemodynamic responses to pump speed modulation, and design feedback control strategies under nominal physiological conditions. However, the selection of a deterministic framework implicitly encodes assumptions about observability, parameter identifiability, and the degree to which patient variability can be neglected. As a result, deterministic models vary significantly in their clinical relevance depending on the application context.

Three major classes of deterministic models dominate the LVAD literature. Time-varying elastance models emphasize ventricular pressure–volume dynamics and provide an intuitive representation of pump–heart interaction \cite{suga1972mathematical, suga1974instantaneous, fernandez2013object, wang2022modeling}. Windkessel-based models abstract the vascular system into lumped resistive and compliant elements, enabling efficient analysis of global hemodynamics and afterload sensitivity \cite{lopez-santana2023computational, mcelroy2020impact, santiago2022design}. Fluid–electro–mechanical models integrate electrical activation, myocardial deformation, and blood flow, offering high-fidelity insight into localized phenomena at the cost of substantial computational complexity \cite{leong2015electromechanics, santiago2018fluid}.

Rather than competing approaches, these modeling classes should be viewed as complementary, each addressing different aspects of LVAD-CVS interaction. Understanding their assumptions and limitations is essential for interpreting simulation outcomes and for selecting appropriate models in control design and clinical translation. The following subsections examine these deterministic approaches in detail.

\subsubsection{Time-Varying Elastance Model:}
\label{Elastance}
The heart can be modeled as an active contractile chamber in which myocardial force generation modulates ventricular volume, producing time-varying pressure responses. The time-varying elastance (TVE) model formalizes this behavior by capturing the pressure–volume (P–V) relationship of the left ventricle (LV) throughout the cardiac cycle and has therefore become one of the most widely used representations in LVAD–cardiovascular system (CVS) modeling \cite{suga1972mathematical}. In this framework, ventricular pressure is governed by the interaction between myocardial contractility, heart rate, and preload-afterload conditions \cite{westerhof2019snapshots, burkhoff2005assessment}, making the model particularly attractive for analyzing ventricular unloading and pump–heart interaction under LVAD support \cite{walley2016left}.

\begin{figure*}[t!]
\centering
\includegraphics[width=0.6\textwidth]{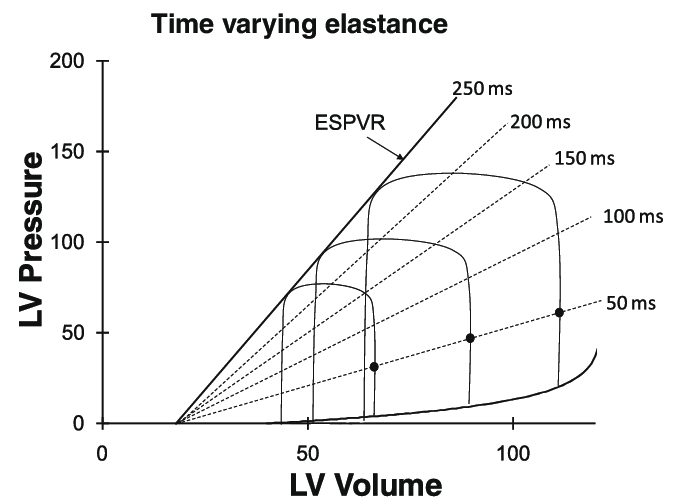}
\caption {End-Systolic Pressure-Volume Relationship (ESPVR) Diagram for three differently loaded cardiac cycles \cite{walley2016left}. }\label{elastance}
\end{figure*}

The physiological origins of elastance theory can be traced to Starling’s observation that increased preload leads to ventricular dilation and changes in chamber stiffness \cite{sarnoff1954ventricular}. Subsequent studies established an approximately linear relationship between stroke volume and end-systolic pressure under controlled conditions \cite{weber1974determinants}. Building on these insights, Suga and colleagues introduced the End-Systolic Pressure-Volume Relationship (ESPVR), demonstrating how variations in preload and afterload shift ventricular P-V trajectories across cardiac cycles \cite{sagawa1977end}. By extending this concept across the entire cardiac cycle, the slope of the P-V relationship, defined as elastance ($\Delta P/\Delta V$), was shown to vary dynamically, giving rise to the concept of time-varying elastance \cite{suga1971theoretical}. Representative ESPVR curves under different loading conditions are illustrated in Figure~\ref{elastance}.

Multiple mathematical formulations of the elastance function have been proposed to capture ventricular dynamics with varying degrees of fidelity. Lankhaar et al.~\cite{lankhaar2009modeling} systematically evaluated six elastance formulations using experimental P-V data, including models with fixed and variable volume intercepts as well as sigmoidal and pressure-dependent elasticity representations. Their results showed that models allowing a varying volume intercept provided superior agreement with measured data based on the Akaike Information Criterion, highlighting the sensitivity of elastance-based models to structural assumptions and parameterization choices.

A widely adopted mathematical representation of time-varying elastance, originally proposed by Suga et al. and subsequently employed in numerous LVAD modeling studies \cite{suga1974instantaneous, son2019stochastic, son2020modelling, simaan2008modeling, tan2016modelling, segers2003systemic, stergiopulos1996determinants, ferreira2007rule}, is given by
\begin{equation}
E(t) = (E_{\text{max}} - E_{\text{min}}),E_n(t_n) + E_{\text{min}},
\label{eq1}
\end{equation}
where $E_n(t_n)$ denotes a normalized “double-hill” function describing the temporal evolution of ventricular stiffness:
\begin{equation}
E_n(t_n) = 1.55 \cdot \left( \frac{t_n}{0.7} \right)^{1.9} \cdot \frac{1}{1 + \left( \frac{t_n}{0.7} \right)^{1.9}} \cdot \frac{1}{1 + \left( \frac{t_n}{1.17} \right)^{21.9}} .
\label{eq2}
\end{equation}
Here, $t_n = t/T_{\text{max}}$ is a dimensionless time variable, $T_{\text{max}} = 0.2 + 0.15,t_c$, and $t_c = 60/\text{HR}$ represents the duration of a single cardiac cycle, with HR denoting heart rate in beats per minute. The parameters $E_{\text{max}}$ and $E_{\text{min}}$ correspond to the end-systolic and end-diastolic pressure–volume relationships (ESPVR and EDPVR), respectively, thereby linking the elastance waveform to clinically meaningful ventricular properties.

Recent extensions of TVE-based models have focused on improving prediction and control of ventricular behavior under continuous-flow LVAD support by incorporating adaptive mechanisms or enhanced coupling with pump dynamics \cite{shi2006numerical, vasudevan2022application, ochsner2017novel, hedayati2024elastance}. In parallel, data-driven approaches, particularly neural network and Long Short-Term Memory (LSTM) models, have been explored to estimate elastance profiles from measurable blood pressure and flow signals, addressing the limited observability of intrinsic ventricular states in implantable systems \cite{tan2024dynamic}.

Despite their physiological interpretability and widespread adoption, time-varying elastance models exhibit fundamental limitations in LVAD applications. The elastance function and its defining parameters are not directly measurable in vivo and typically require offline calibration or simplifying assumptions, reducing robustness under patient-specific variability and disease progression. Moreover, the assumption of quasi-periodic ventricular behavior becomes increasingly tenuous under continuous-flow LVAD support, where altered loading conditions, suction events, and reduced pulsatility can invalidate classical elastance relationships. Consequently, while TVE models provide valuable insight into ventricular mechanics and pump–heart interaction, they are often insufficient as standalone representations for real-time LVAD control and must be complemented by alternative or hybrid modeling approaches.

\subsubsection{Lumped-Parameter Models}

Windkessel models provide a lumped-parameter representation of the arterial system based on an electrical–hydraulic analogy, in which blood pressure is treated as voltage and blood flow as current. Originally formalized by Frank in 1899, the classical Windkessel framework represents the arterial tree using combinations of resistance, compliance, and inertance elements to capture the dominant features of arterial load \cite{frank1899erste}. By neglecting spatial variations such as vessel length, wave propagation, and velocity profiles, Windkessel models are classified as zero-dimensional (0D) models, treating the arterial system as a single or multiple compartment.

The primary appeal of Windkessel models in LVAD research lies in their simplicity and computational efficiency. These models enable rapid simulation of ventriculo–arterial coupling and are widely used to construct hydraulic boundary conditions for isolated hearts, mock circulatory loops, and assist devices \cite{westerhof1969analog, ferrari2005development, ferrari2009role, luo2007using}. However, this simplicity comes at the cost of reduced physiological detail, particularly in representing wave reflections, spatial pressure gradients, and distal perfusion dynamics.

\paragraph{2.1.2.1  Single-Compartment (Windkessel) Models}

Single-compartment Windkessel models approximate the arterial system as a single lumped load and are commonly categorized according to the number of elements employed.

\begin{itemize}
\item \textbf{Two-Element Windkessel Model (RC):}  
The simplest formulation incorporates arterial compliance ($C$) and peripheral resistance ($R$), capturing the exponential decay of arterial pressure during diastole. While computationally efficient, the two-element model fails to represent systolic dynamics and characteristic impedance effects \cite{frank1899erste}.

\item \textbf{Three-Element Windkessel Model (RCR):}  
By introducing a characteristic impedance ($R_c$) in series with the compliance–resistance pair, the three-element model improves representation of systolic pressure and flow dynamics \cite{westerhof1971artificial}. Despite this improvement, it may overestimate arterial compliance and underestimate characteristic impedance under certain conditions \cite{stergiopulos1999total}.

\item \textbf{Four-Element Windkessel Models:}  
To account for blood flow inertia, a fourth element representing inertance ($L$) is added either in series or parallel with the existing RC components \cite{stergiopulos1999total, grant1987characterization}. These models improve parameter estimation accuracy and frequency-domain behavior but introduce additional complexity.

\item \textbf{Viscoelastic and Inertance-Viscoelastic Models:}  
Viscoelastic Windkessel (VW) models incorporate vessel wall viscoelasticity through a Voigt cell, enhancing compliance estimation under dynamic loading \cite{burattini1998complex}. The inertance–viscoelastic Windkessel (IVW) model further integrates inertial effects, improving representation of blood inertia and vessel wall viscosity \cite{burattini2007development}.

\begin{figure*}[t!]
  \centering
  \begin{subfigure}[b]{0.45\textwidth}
        \centering
        \includegraphics[width=\textwidth]{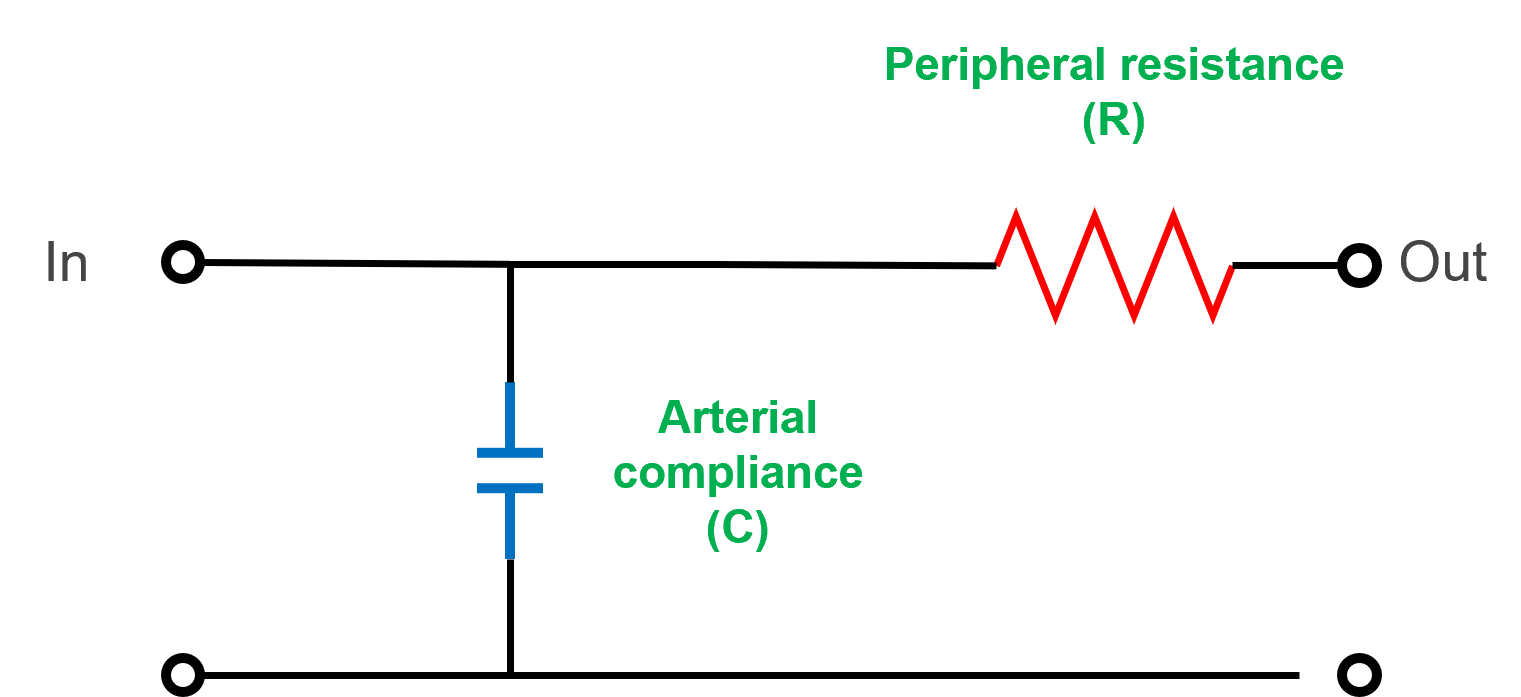}
        \caption{2-element Windkessel}
        \label{WK2}
    \end{subfigure}
    \hfill
    \begin{subfigure}[b]{0.45\textwidth}
        \centering
        \includegraphics[width=\textwidth]{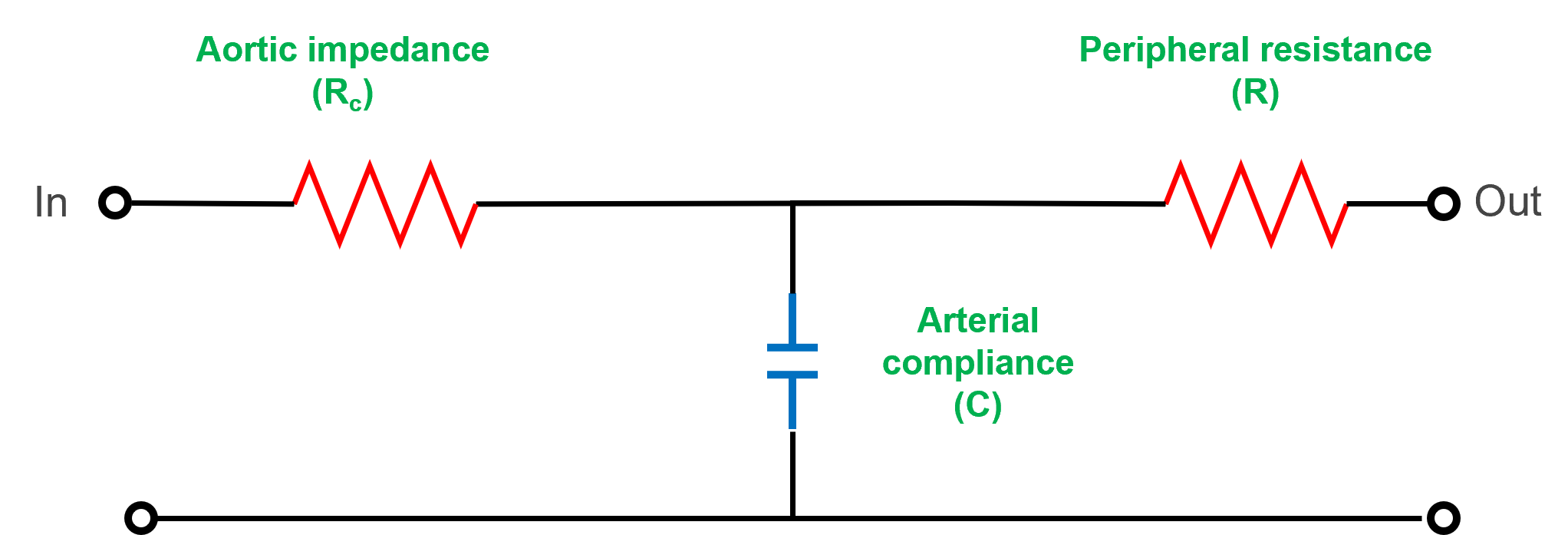}
        \caption{3-element Windkessel}
        \label{WK3}
    \end{subfigure}
    \hfill
    \begin{subfigure}[b]{0.45\textwidth}
        \centering
        \includegraphics[width=\textwidth]{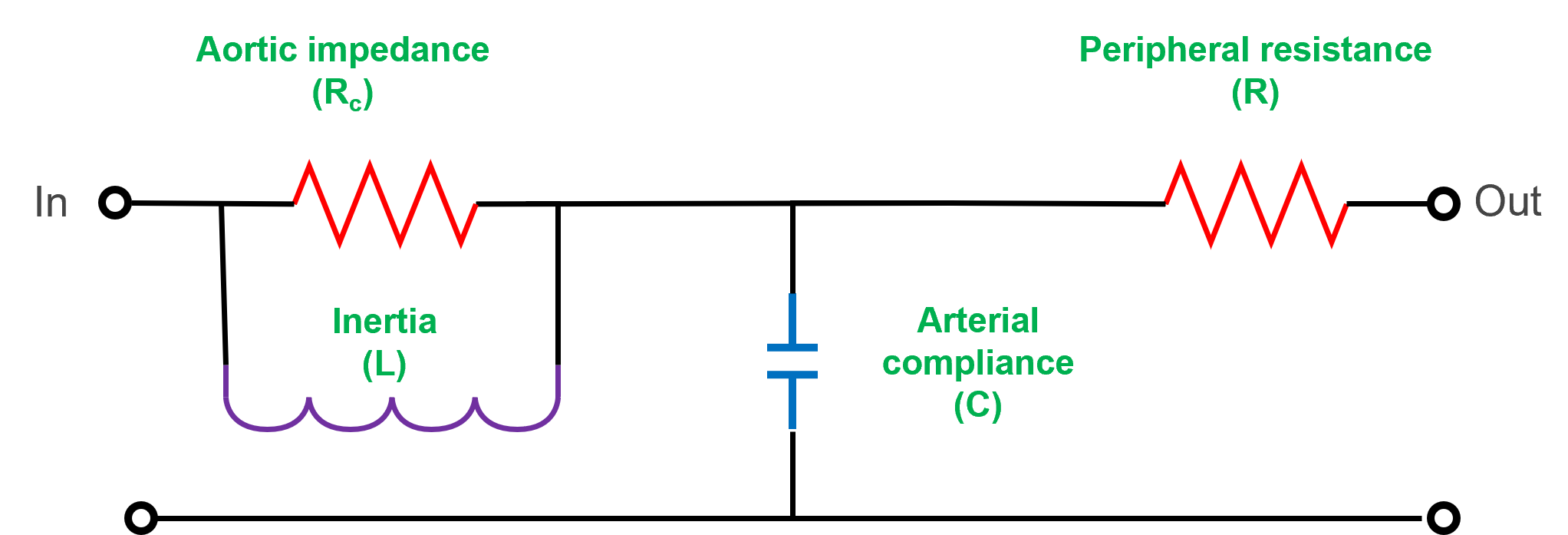}
        \caption{4-element Windkessel (W4P)}
        \label{W4P}
    \end{subfigure}
    \hfill
    \begin{subfigure}[b]{0.45\textwidth}
        \centering
        \includegraphics[width=\textwidth]{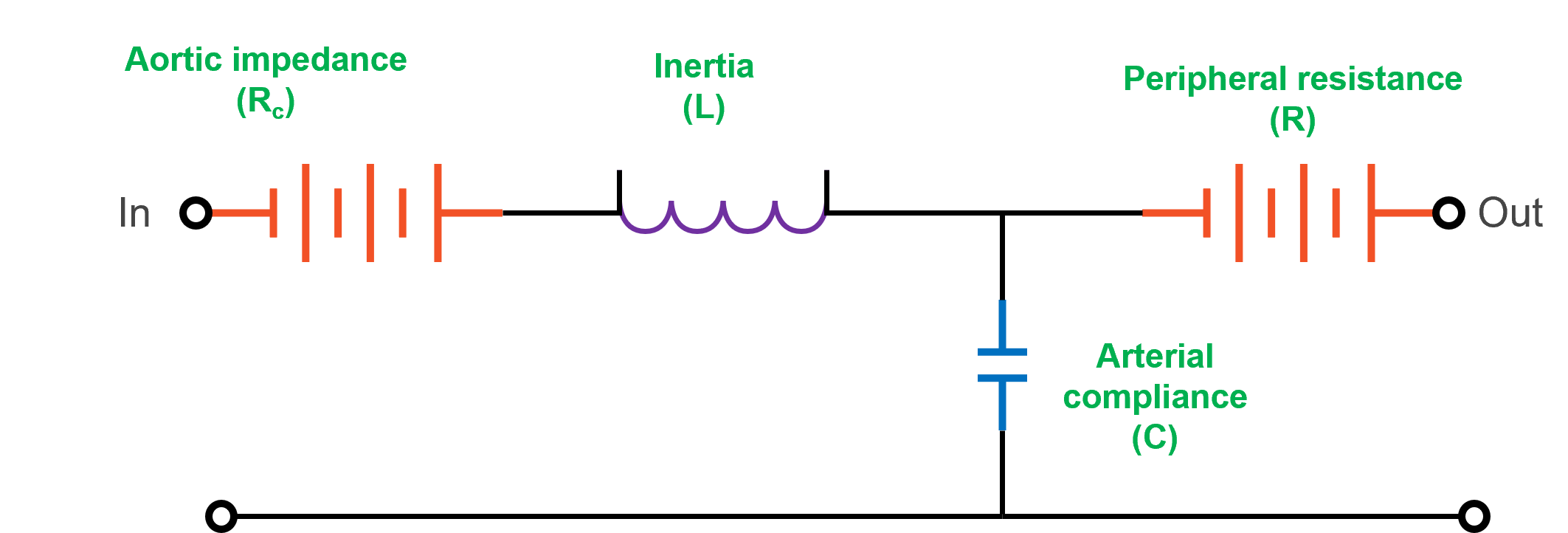}
        \caption{4-element Windkessel (W4S)}
        \label{W4S}
    \end{subfigure}
    \hfill
    \begin{subfigure}[b]{0.45\textwidth}
        \centering
        \includegraphics[width=\textwidth]{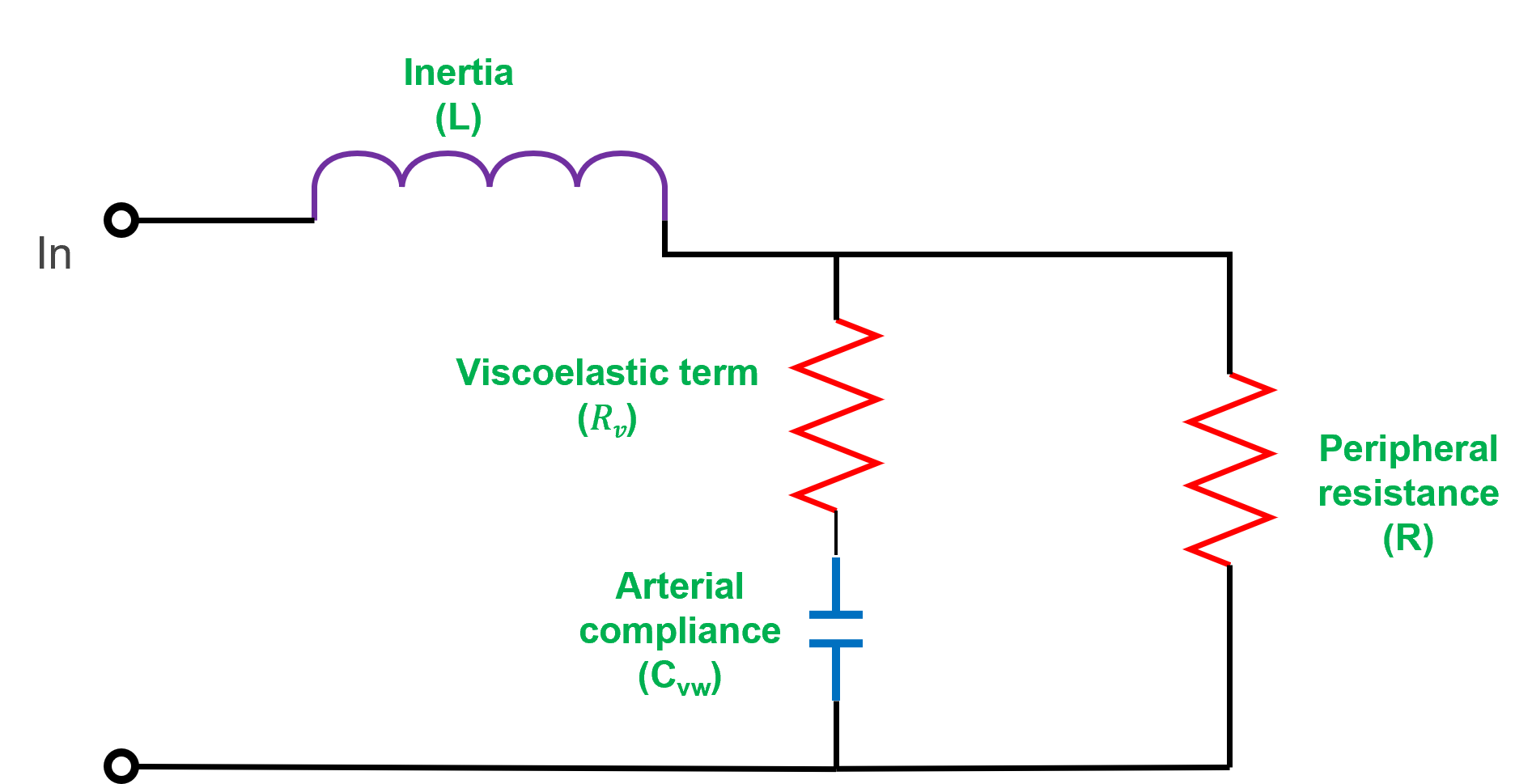}
        \caption{Inertance Viscoelastic Windkessel (IVW)}
        \label{IVW}
    \end{subfigure}
    \hfill
    \begin{subfigure}[b]{0.45\textwidth}
        \centering
        \includegraphics[width=\textwidth]{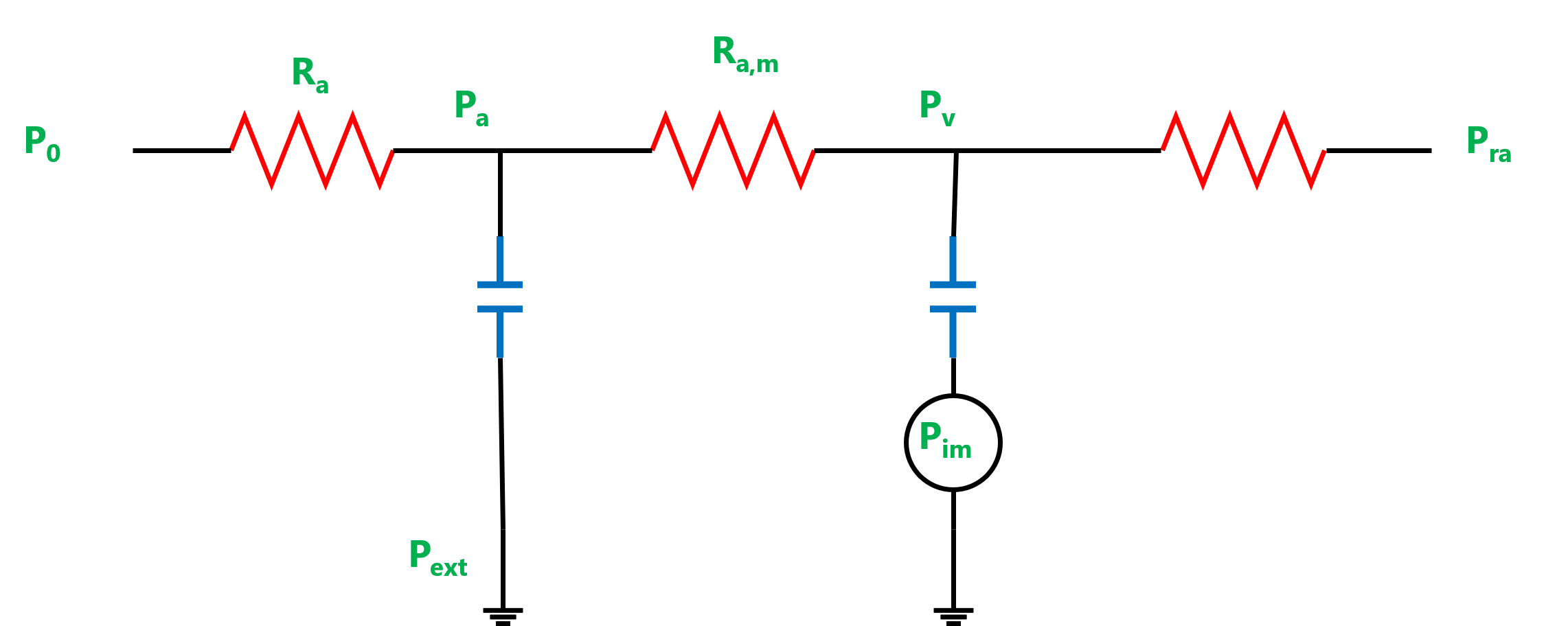}
        \caption{5-element Windkessel}
        \label{WK5}
    \end{subfigure}
  \caption{Comparison of Windkessel models \cite{frank1899erste, westerhof1971artificial, stergiopulos1999total, grant1987characterization, burattini2007development, jonavsova2021relevance, fernandes2023modeling}.}
  \label{fig:WK_models}
\end{figure*}

\FloatBarrier

\item \textbf{Five-Element Windkessel Model:}  
The five-element model separates arterial compliance into proximal and distal components and introduces additional resistive elements to represent arterial and venous segments. This formulation offers improved physiological realism and more accurate impedance matching across a broader frequency range \cite{noordergraaf2012circulatory, jonavsova2021relevance}. Recent studies have integrated five-element Windkessel models with systolic pulse transit time (sPTT) approaches in commercial solvers such as ANSYS to enhance coronary flow and fractional flow reserve (FFR) estimation \cite{fernandes2023modeling}.
\end{itemize}

Despite their utility, single-compartment Windkessel models cannot capture wave propagation, reflection phenomena, or spatial pressure variations along the arterial tree. These limitations restrict their ability to accurately predict distal pressure conditions and local hemodynamics, particularly under LVAD-induced continuous-flow operation.

To address these shortcomings, several extensions have been proposed. Fractional-order Windkessel models introduce non-integer order elements to better represent arterial viscoelasticity and frequency-dependent behavior \cite{bahloul2019fractional, bahloul2023fractional}. Hybrid formulations combining Windkessel models with pulse transit time or solitonic wave representations have further improved pressure prediction and vascular aging characterization \cite{ujiie2022solitonic, fernandes2023modeling}.

\paragraph{2.1.2.2  Multiple-Compartment LPMs}

While Windkessel models provide global arterial load approximation, they are insufficient for capturing distributed cardiovascular dynamics. Lumped-parameter models (LPMs) extend the Windkessel concept by representing the CVS as a network of resistive, compliant, and inertial elements or elastic chambers, enabling multi-compartment analysis of pressure–flow interactions \cite{abdi2015lumped, ferreira2005nonlinear}. The voltage–pressure and current–flow analogy allows efficient simulation of ventricular–arterial coupling and device–circulation interaction across multiple physiological states.

Lower-order LPMs prioritize simplicity and control-oriented modeling. A fourth-order LPM represents the major CVS components using time-varying compliance to model cardiac pumping and diodes to represent valve behavior, typically assuming normal right ventricular function \cite{ferreira2005dynamical, simaan2011left}.

\begin{figure*}[!b]
  \centering
  \includegraphics[width=0.6\textwidth]{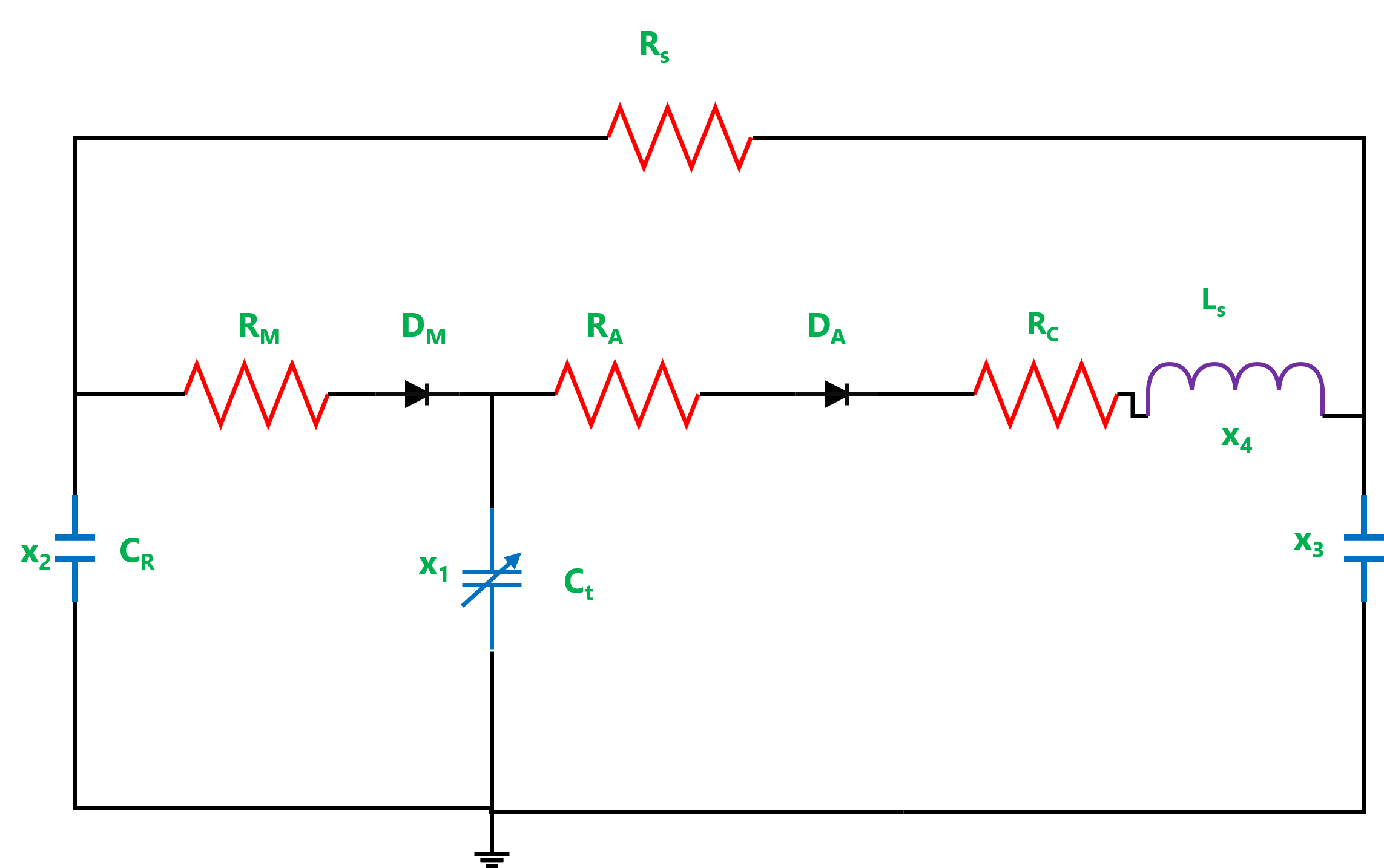}
  \caption{$4^{th}$ order lumped-parameter model of the CVS.}
  \label{4th LPM}
\end{figure*}

Higher-order LPMs progressively incorporate additional physiological detail. Fifth-order models introduce aortic compliance or LVAD dynamics to improve arterial waveform fidelity and capture pump–heart interaction \cite{ferreira2005nonlinear, schima1990computer}.

\begin{figure*}[t!]
  \centering
  \begin{subfigure}[b]{0.7\textwidth}
        \centering
        \includegraphics[width=\textwidth]{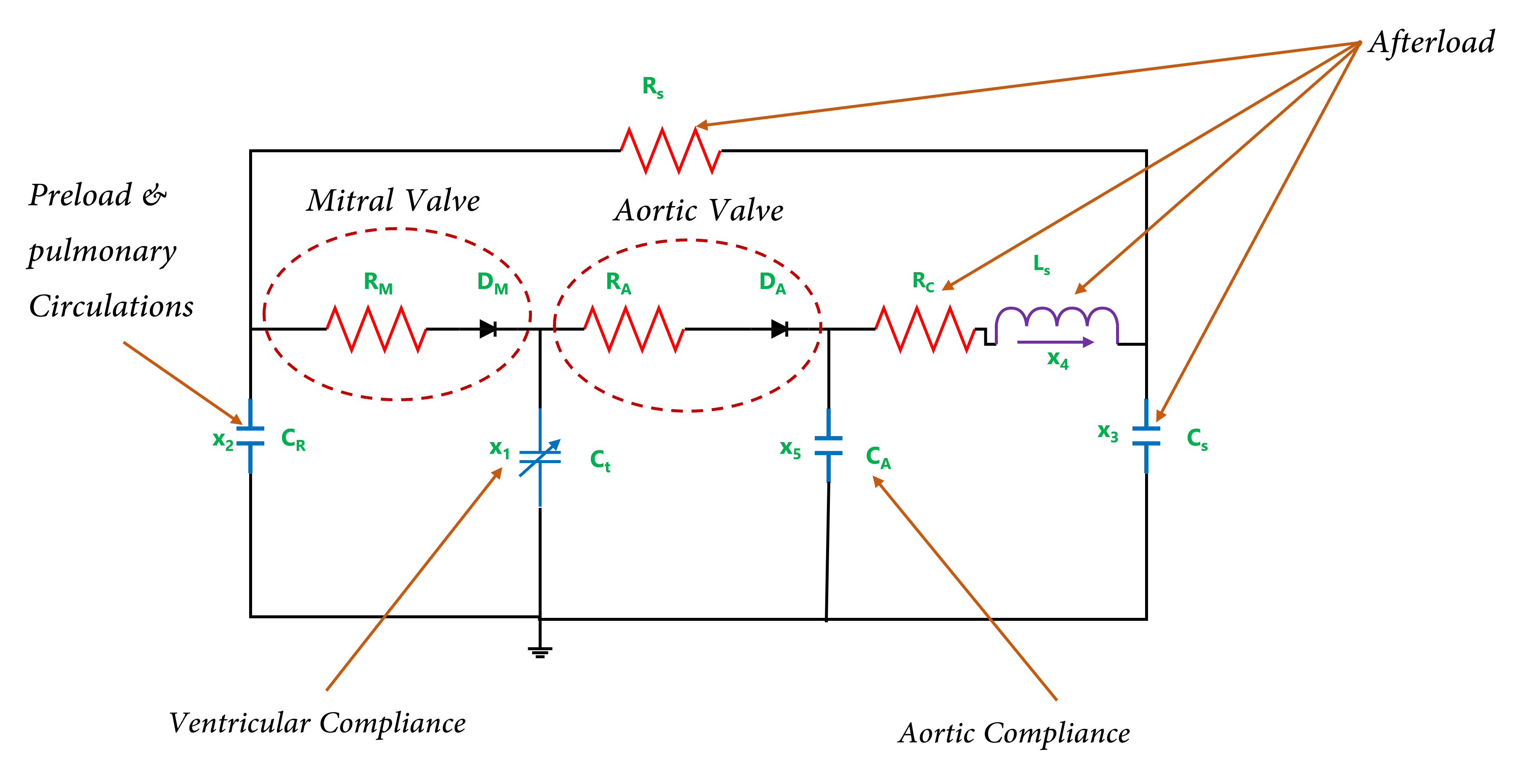}
        \caption{$5^{th}$ order CVS without LVAD}
        \label{5th LMP2}
    \end{subfigure}
    \hfill
    \begin{subfigure}[b]{0.45\textwidth}
        \centering
        \includegraphics[width=\textwidth]{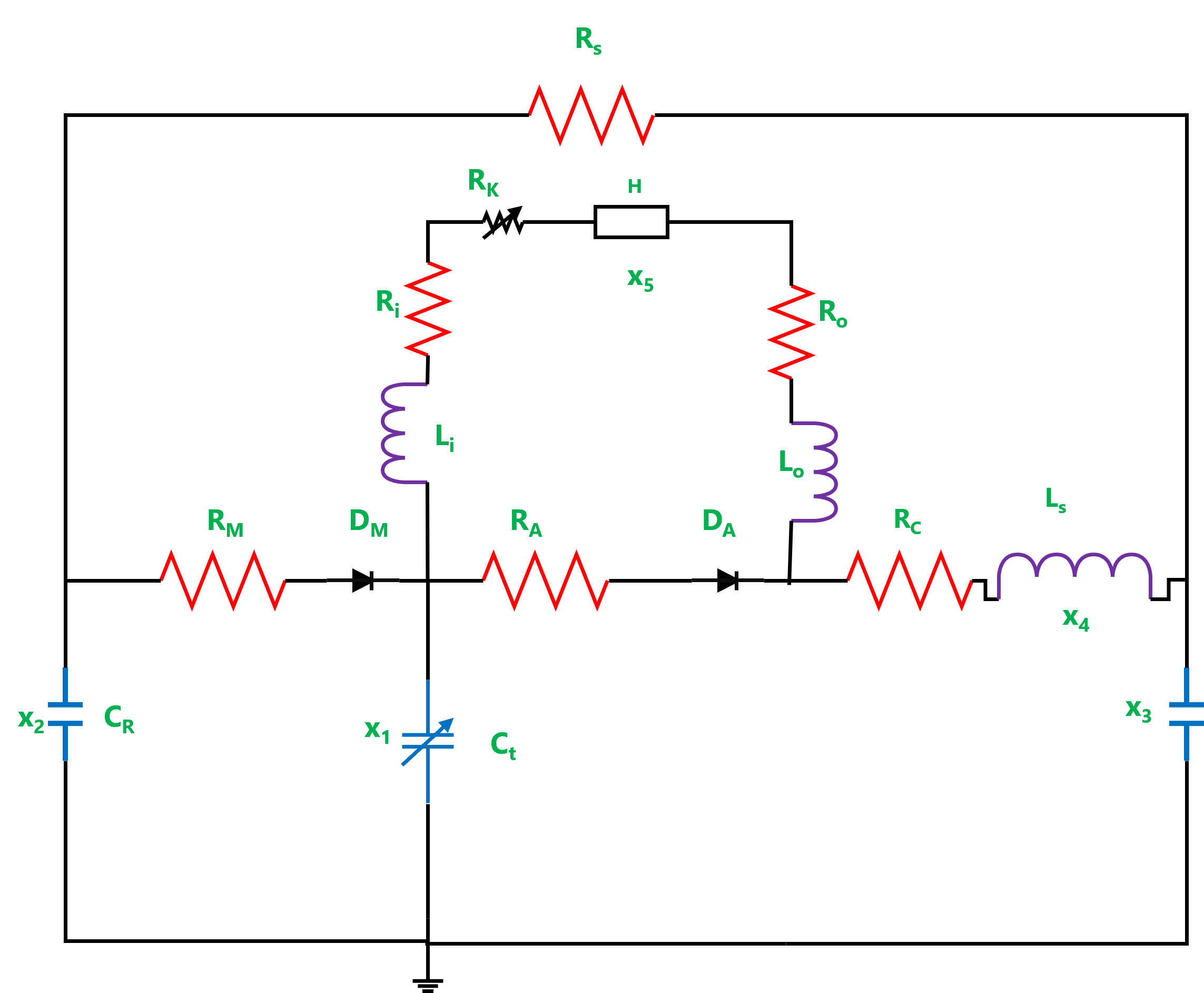}
        \caption{$5^{th}$ order CVS with LVAD}
        \label{5th LMP}
    \end{subfigure}
    \caption{Comparison of fifth-order lumped-parameter models \cite{ferreira2005nonlinear}.}
    \label{fig:5th_LPM}
\end{figure*}

Sixth-order and higher models further enhance physiological realism by incorporating both aortic compliance and LVAD flow states, enabling more accurate representation of systemic circulation and ventricular–arterial coupling \cite{simaan2008dynamical}.

\begin{figure*}[t!]
\centering
\includegraphics[width=0.5\textwidth]{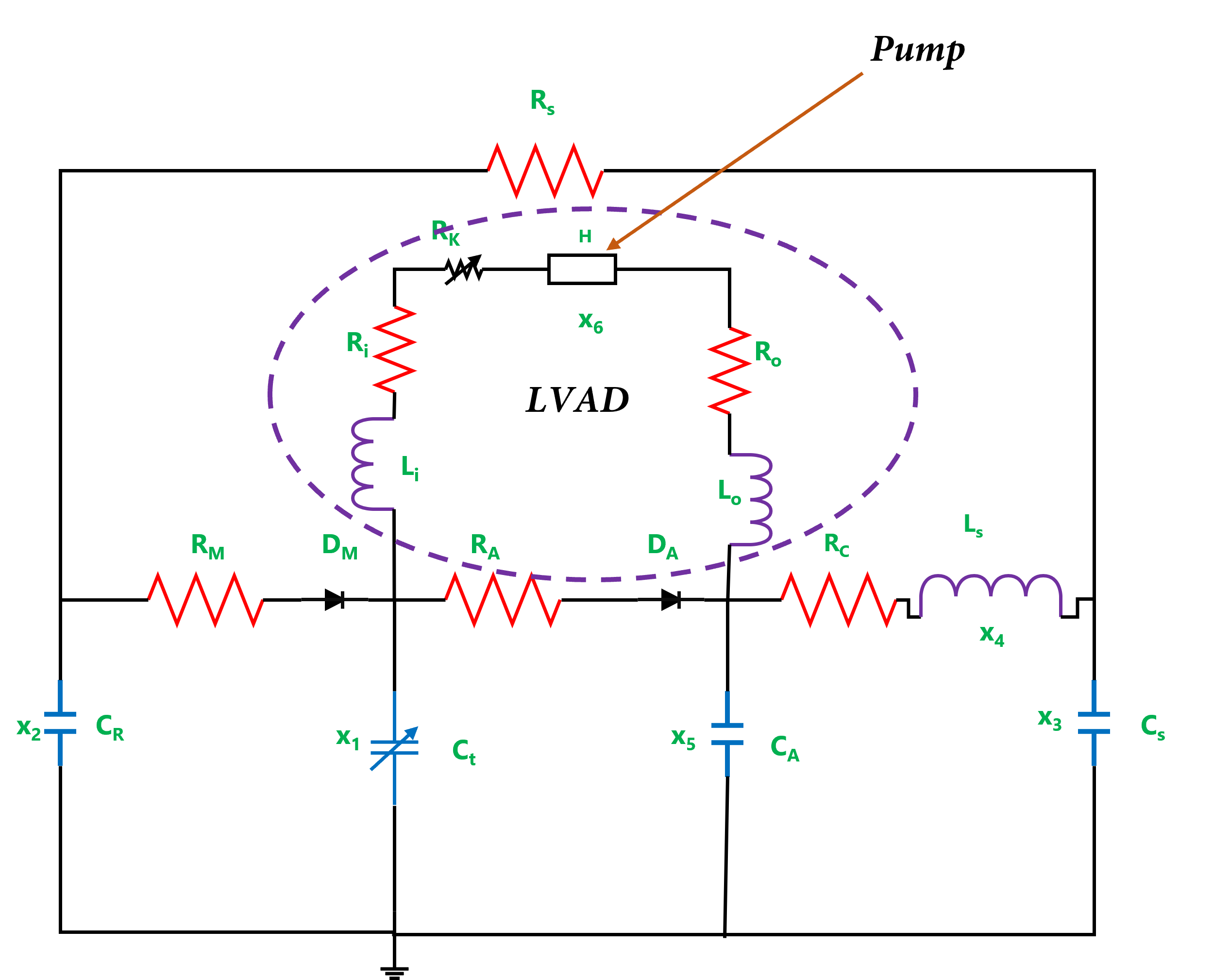}
\caption{Cardiovascular pump $6^{th}$ order model \cite{simaan2008dynamical}.}
\label{6th LMP}
\end{figure*}

\begin{figure*}[t!]
\centering
\includegraphics[width=0.8\textwidth]{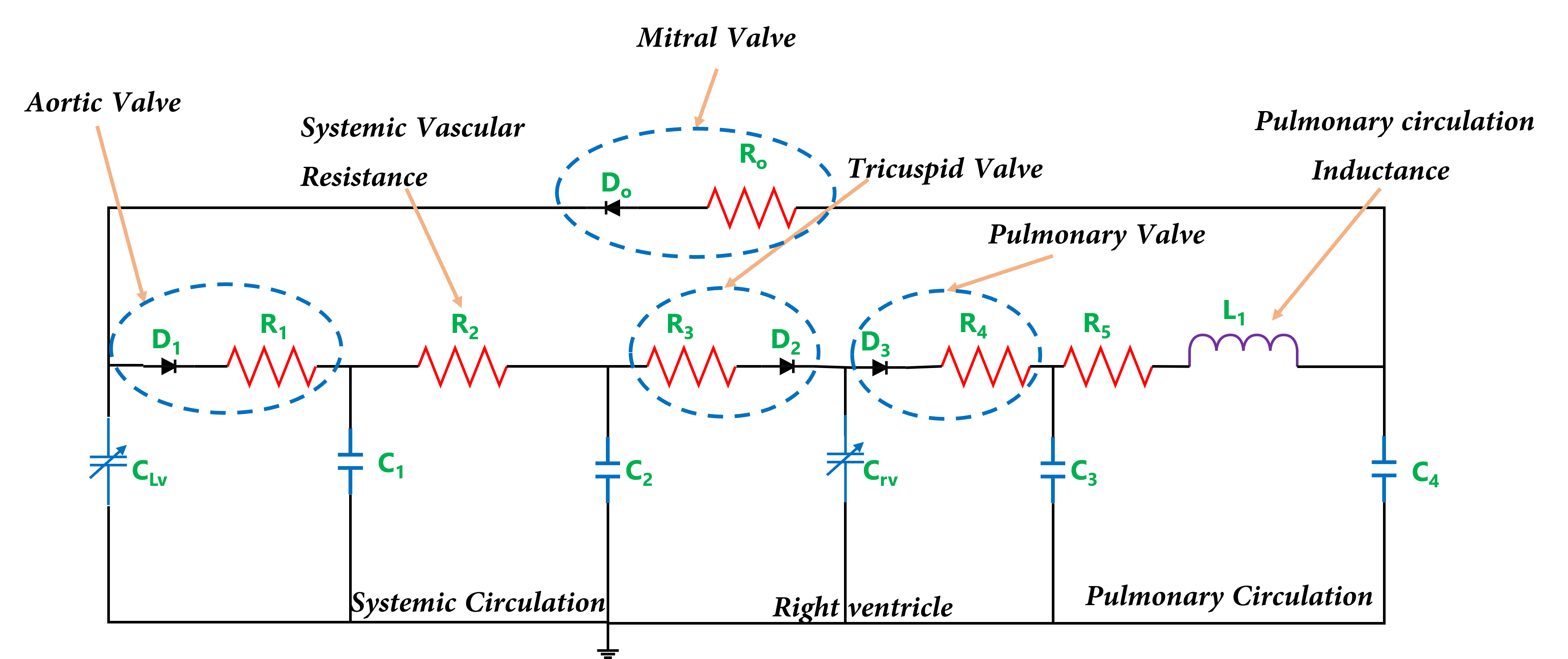}
\caption{Biventricular cardiovascular $7^{th}$ order lumped-parameter model.}
\label{7th LMP}
\end{figure*}

Advanced LPMs integrate biventricular dynamics, baroreflex regulation, exercise adaptation, and patient-specific parameterization. Examples include the Utah Circulation Model \cite{giridharan2002modeling}, baroreflex-enabled LVAD control models \cite{son2020modelling, rubtsova2021lumped}, and disease-specific formulations addressing valve disorders, congenital defects, and heart failure progression \cite{ortiz2022dynamic, liu2024development}. Hybrid LPM-CFD approaches have also been employed to provide realistic boundary conditions for high-fidelity flow simulations \cite{aka2022complete, garven2025pediatric}.

\FloatBarrier

\subsubsection{Fluid-Electro-Mechanical Model:}

\begin{figure*}[t!]
  \centering
  \includegraphics[width=0.8\textwidth]{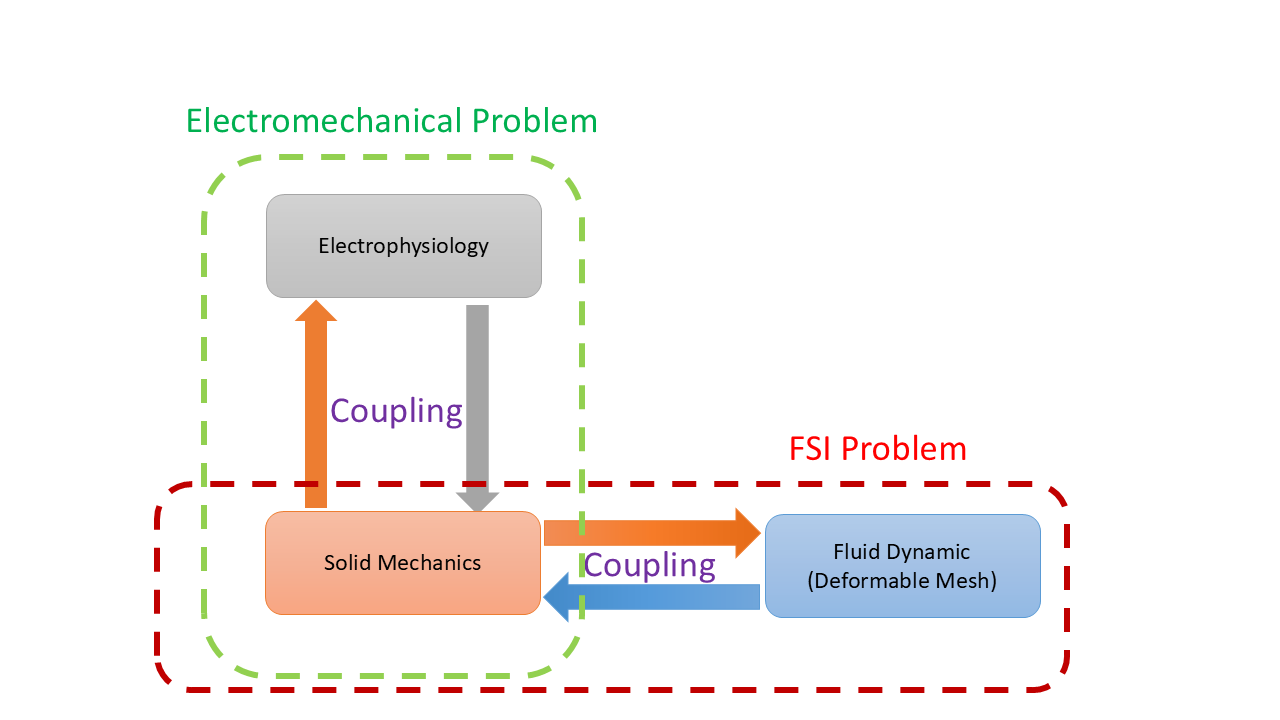}
  \caption {Schematic representation of a fluid-electro-mechanical cardiac model. Electrical activation (electrophysiology) is coupled with myocardial deformation through electromechanical interaction, while ventricular wall motion is bidirectionally coupled with blood flow via a fluid-structure interaction (FSI) framework. \cite{leong2015electromechanics}.}\label{electro_mechanical}
\end{figure*}

As the name indicates, the Fluid-Electro-Mechanical model is a computational framework that integrates the electrical, mechanical, and fluidic aspects of cardiac function and LVAD operation. As illustrated in Fig.~\ref{electro_mechanical}, the model is naturally decomposed into an electromechanical problem, coupling electrophysiology with myocardial solid mechanics, and a fluid-structure interaction (FSI) problem, coupling ventricular wall deformation with intracardiac blood flow. This modeling approach is motivated by the fact that the heartbeat can be described as the interaction of three fundamental processes: (a) electrical depolarization, (b) myocardial mechanics, and (c) blood flow dynamics \cite{leong2015electromechanics}.

In FEM formulations, electrical propagation is typically modeled using continuum reaction-diffusion equations, myocardial mechanics are described using finite elasticity theory, and blood flow is governed by the Navier-Stokes equations in deformable domains.

In the context of LVAD-supported circulation, FEM models enable detailed investigation of pump-heart interactions, including myocardial stress distribution, ventricular unloading patterns, suction phenomena, and electromechanical dyssynchrony induced by continuous-flow assistance. However, despite their high physiological fidelity, FEM models require detailed patient-specific geometry, extensive parameterization, and substantial computational resources. Consequently, their application is largely confined to offline analysis, device design, and mechanistic studies rather than real-time control or routine clinical deployment.

\begin{table*}[!t]
\centering
\caption{Comparison of commonly used deterministic cardiovascular models for LVAD-supported circulation.}
\label{tab:deterministic_models}
\begin{tabular}{p{2.5cm} p{3cm} p{3cm} p{2.5cm} p{2.5cm}}
\hline
\textbf{Model Type} & \textbf{Modeling Framework} & \textbf{Physiological Representation} & \textbf{Advantages} & \textbf{Limitations} \\
\hline

\textbf{Time-Varying Elastance Model} & 
Time-dependent ventricular elastance function describing the pressure–volume relationship of the left ventricle using ordinary differential equations. &
Captures ventricular contraction dynamics, preload/afterload interaction, and pressure–volume loops of the cardiac cycle. &
Computationally efficient, widely validated, suitable for control design and system-level simulations. &
Limited spatial representation of cardiac mechanics; assumes simplified ventricular geometry and may not fully capture LVAD–heart interaction under varying physiological conditions. \\

\hline

\textbf{Windkessel Models} &
Lumped-parameter representation of the arterial system using resistance, compliance, and inertance elements derived from electrical analogies. &
Represents global arterial load and ventricular–arterial coupling through simplified vascular dynamics. &
Simple implementation, low computational cost, useful for studying systemic hemodynamics and designing boundary conditions. &
Cannot capture spatial flow distribution, wave propagation, or detailed vascular geometry. \\

\hline

\textbf{Lumped Parameter Models (Multi-Compartment)} &
Network-based extension of Windkessel models representing the cardiovascular system using multiple RLC elements and elastic chambers. &
Captures interactions between cardiac chambers, systemic and pulmonary circulation, and LVAD dynamics. &
Provides more detailed pressure–flow relationships and improved physiological realism compared to basic Windkessel models. &
Requires larger parameter sets and calibration; still limited in capturing detailed fluid dynamics. \\

\hline

\textbf{Fluid-Electro Mechanical Models} &
Multiphysics framework coupling electrophysiology, myocardial mechanics, and fluid dynamics using finite element methods and Navier–Stokes equations. &
Represents electrical activation, myocardial contraction, and blood flow within deformable cardiac chambers. &
High physiological fidelity; enables detailed simulation of cardiac electromechanics and LVAD–heart interaction. &
Computationally expensive, requires extensive patient-specific data and complex numerical implementation. \\

\hline
\end{tabular}
\end{table*}

Table~\ref{tab:deterministic_models} highlights the trade-offs among deterministic modeling approaches used for LVAD-supported circulation. While these models provide valuable physiological insight and are widely used for simulation and control design, they rely on fixed parameters and deterministic assumptions that limit their ability to capture patient-specific variability, measurement uncertainty, and stochastic physiological fluctuations. These limitations are particularly significant in clinical settings, where cardiovascular dynamics exhibit inherent variability and are influenced by unmeasured disturbances. To address these challenges, stochastic modeling approaches have been proposed to explicitly incorporate uncertainty and improve robustness in LVAD modeling and control.

\subsection{Stochastic Models}
\label{Stochastic}

Deterministic cardiovascular models, while physiologically interpretable, assume fixed parameters and known system dynamics. However, cardiovascular behavior in LVAD-supported patients is inherently variable and influenced by patient-specific factors, measurement noise, and unmodeled physiological disturbances. Ignoring such variability may lead to suboptimal control performance and, in extreme cases, adverse clinical outcomes \cite{son2019}. Consequently, stochastic modeling approaches have been introduced to explicitly incorporate uncertainty and improve the robustness of LVAD modeling and control frameworks.

In the context of LVAD-supported circulation, uncertainty arises from multiple sources, including inter-patient variability in cardiovascular parameters, intra-patient physiological fluctuations due to activity or disease progression, and limitations in sensor accuracy. To address these challenges, several stochastic modeling techniques have been proposed.

Monte Carlo sampling is a widely used approach for uncertainty quantification, enabling probabilistic analysis of model outputs under parameter variability \cite{spanos1998monte}. However, its practical application in LVAD systems is limited by high computational cost, as a large number of simulations are required to achieve statistical convergence.

More recently, data-driven stochastic approaches have been explored. For instance, Gaussian Process models have been employed to estimate hematocrit (HCT) levels in LVAD patients using data obtained from hybrid mock circulation systems \cite{petrou2018viscosity}. These models provide probabilistic predictions and uncertainty bounds, making them suitable for clinical monitoring applications.

Another promising approach is the generalized polynomial chaos (gPC) expansion, which enables efficient representation of stochastic dynamics within a state-space framework \cite{son2019}. This method has been applied to model variations in pump flow and to design self-tuning feedback controllers capable of adapting to changing physiological conditions, such as varying levels of physical activity.

Despite these advancements, stochastic models face several challenges, including increased model complexity, difficulties in parameter identification, and limited clinical validation. Furthermore, the integration of stochastic models into real-time LVAD control systems remains an open research problem, particularly due to computational constraints and the lack of reliable implantable sensors for continuous monitoring.

Overall, stochastic modeling provides a critical extension to deterministic approaches by enabling the explicit treatment of uncertainty. However, further research is required to bridge the gap between theoretical developments and clinically deployable LVAD control systems.

\subsection{Hybrid Modeling Perspectives}
\label{hybrid_modeling}

While deterministic and stochastic models provide complementary insights into cardiovascular dynamics, neither framework alone fully captures the complexity of LVAD-supported physiology. Deterministic models offer interpretability and computational efficiency but are limited in handling uncertainty and patient-specific variability. Conversely, stochastic models incorporate uncertainty but often introduce increased computational complexity and challenges in parameter identification.

To address these limitations, recent research has explored hybrid modeling strategies that combine physics-based representations with data-driven components. In this context, machine learning techniques are not used as standalone replacements for physiological models, but rather as complementary tools to enhance model adaptability and accuracy. For example, data-driven modules can be integrated with lumped-parameter or elastance-based models to estimate uncertain parameters, compensate for modeling inaccuracies, or enable patient-specific calibration.

Such hybrid approaches provide a balance between physiological interpretability and predictive flexibility, making them particularly suitable for LVAD applications where both accuracy and robustness are critical. However, challenges remain in terms of model validation, data availability, and integration into real-time control frameworks. These aspects are further discussed in the context of control strategies in Section~\ref{Control}.

\begin{table*}[!t]
\centering
\caption{Comparative synthesis of LVAD modeling paradigms highlighting their ability to handle uncertainty, patient variability, and clinical translation challenges.}
\label{tab:modeling_paradigms}

\begin{tabular}{|p{2cm}|p{2cm}|p{2.5cm}|p{2.5cm}|p{2cm}|p{2.8cm}|}
\hline
\textbf{Modeling Paradigm} & \textbf{Uncertainty Handling} & \textbf{Patient-Specific Adaptability} & \textbf{Computational Cost} & \textbf{Real-Time Suitability} & \textbf{Key Limitation / Research Gap} \\
\hline

\textbf{Deterministic Models} &
Low (fixed parameters) &
Limited &
Low &
High &
Inability to capture physiological variability and uncertainty; limited personalization \\

\hline

\textbf{Stochastic Models} &
High (probabilistic frameworks such as Monte Carlo, gPC) &
Moderate-High &
High &
Low &
Computational burden and difficulty in real-time implementation; limited clinical validation \\

\hline

\textbf{Hybrid Models (Physics + Data)} &
Moderate-High &
High &
Moderate &
Moderate &
Integration complexity, need for large datasets, and lack of standardized validation frameworks \\

\hline
\end{tabular}
\end{table*}

Table~\ref{tab:modeling_paradigms} reveals a fundamental trade-off in LVAD modeling approaches. Deterministic models remain suitable for real-time implementation but lack the ability to capture uncertainty and patient-specific variability. Stochastic models improve robustness by incorporating uncertainty but introduce significant computational complexity, limiting their clinical applicability. Hybrid modeling approaches emerge as a promising direction by combining physiological interpretability with data-driven adaptability. However, their successful translation into clinical practice requires overcoming challenges related to data availability, model validation, and real-time deployment.

\section{Control Strategies}
\label{Control}

While significant progress has been made in the modeling of LVAD-supported cardiovascular systems, the ultimate clinical objective is to translate these models into effective real-time control strategies. LVADs have demonstrated considerable success in improving survival in patients with advanced heart failure; however, most commercially available devices operate at a constant speed, manually adjusted by clinicians. Although this approach is sufficient for short-term support, it is inadequate for long-term therapy, where physiological demands vary continuously due to changes in activity level, posture, and emotional state \cite{li2022intelligent, wang2015rotary}.

The inability of fixed-speed operation to adapt to dynamic physiological conditions may lead to adverse events, including ventricular suction, myocardial injury, arrhythmias, and suboptimal perfusion \cite{liu2025multi, cysyk2019cannula}. Consequently, there is a critical need for automatic control systems capable of regulating pump speed in real time to maintain hemodynamic stability and ensure patient safety.

At the core of LVAD control lies the concept of feedback regulation, in which a physiological variable—either directly measured or estimated—is used to adjust pump operation. This process aims to emulate the natural autoregulatory mechanisms of the cardiovascular system by comparing the measured signal to a desired reference state \cite{schmid2020pathophysiological}. Various control targets have been proposed in the literature, including pressure difference \cite{li2022intelligent}, inlet flow rate \cite{bullister2002physiologic}, pump pressure \cite{casas2007minimal}, end-diastolic pressure \cite{fetanat2019physiological}, and Starling-like control mechanisms \cite{salamonsen2012theoretical}.

Despite these advances, the design of robust LVAD controllers remains challenging due to nonlinear cardiovascular dynamics, inter- and intra-patient variability, and the limited availability of reliable implantable sensors. These constraints create a fundamental trade-off between physiological accuracy, robustness, and practical implementability. As a result, a wide range of control strategies has been developed, spanning from classical linear controllers to advanced adaptive and data-driven approaches.

In the following subsections, these control methodologies are reviewed in a progressive manner, highlighting their underlying principles, advantages, limitations, and suitability for clinical translation.

\subsection{Classical Control: Proportional–Integral–Derivative (PID)}
\label{PID}

Proportional–Integral–Derivative (PID) control represents the most widely adopted classical control strategy in LVAD systems due to its simplicity, ease of implementation, and low computational cost. In this framework, the pump speed ($\omega$) is regulated by minimizing the error between a desired physiological reference signal (e.g., pressure difference $\Delta P_r$) and the measured or estimated output ($\Delta P$). The control objective is to maintain hemodynamic stability while preventing adverse events such as ventricular suction, thrombus formation, or right ventricular dysfunction \cite{tanaka2006detection, vollkron2004development, reesink2007suction}.

Early implementations demonstrated the feasibility of PID-based control in LVAD applications. For instance, Parnis et al. employed a proportional controller in the Jarvik 2000 VAD and validated its performance in long-term animal studies \cite{parnis1997progress}. Similarly, Waters et al. proposed a proportional–integral (PI) controller to maintain physiological perfusion under varying activity levels \cite{waters1999motor}. More advanced formulations incorporate physiological principles, such as Frank–Starling-based control using left atrial pressure as feedback \cite{stevens2011frank}, as well as strategies aimed at restoring pulsatility in continuous-flow devices \cite{ramesh2021considerations}.

Despite these advantages, the applicability of PID controllers in LVAD systems is fundamentally limited by the highly nonlinear and time-varying nature of cardiovascular dynamics. Fixed gain parameters are typically unable to accommodate inter- and intra-patient variability, leading to degraded performance under changing physiological conditions. Furthermore, many PID-based implementations rely on direct measurements of pressure or flow, which are difficult to obtain reliably in vivo due to the lack of durable implantable sensors and the associated risks of thrombosis and infection \cite{kitamura2000physical}.

In addition, simplified modeling assumptions commonly used in PID design—such as steady-state flow conditions, neglect of valve dynamics, and reduced-order pump representations—limit physiological accuracy and may result in suboptimal control performance. These limitations highlight the need for more advanced control strategies capable of handling uncertainty, nonlinearity, and patient-specific variability, thereby motivating the development of intelligent, robust, and adaptive control approaches discussed in subsequent sections.

\subsection{Intelligent Control: Fuzzy Logic Control (FLC)}
\label{FLC}

Fuzzy Logic Control (FLC) has emerged as an effective alternative to classical control strategies in LVAD systems, particularly to address the limitations of PID controllers in handling nonlinear and uncertain cardiovascular dynamics. Unlike model-based approaches, FLC does not require an explicit mathematical representation of the system, making it well-suited for applications characterized by patient variability and incomplete physiological knowledge.

In LVAD applications, FLC schemes utilize linguistic rules and expert knowledge to map input variables—such as pressure, flow, or pump speed—into appropriate control actions. This flexibility enables FLC to adapt to varying physiological conditions, including changes in patient activity levels and hemodynamic states. Various implementations have demonstrated its utility in tasks such as failure detection \cite{yoshizawa1994assessing}, sensorless flow regulation \cite{kaufmann1995fuzzy, kaufmann1997implantable}, real-time control on FPGA platforms \cite{santos2019intelligent}, and detection of preload and afterload variations \cite{alomari2012developments}.

To further enhance performance, hybrid control schemes combining FLC with classical controllers have been proposed. For example, FLC–PI controllers have been used to maintain safe flow rates and prevent ventricular suction by leveraging the robustness of fuzzy inference alongside the steady-state accuracy of PI control \cite{choi2001sensorless}. More recently, FLC has been integrated with data-driven approaches, such as deep reinforcement learning (DRL), to improve adaptability and responsiveness under dynamic physiological conditions \cite{li2022intelligent, azizkhani2022supervised, wang2024aortic}.

Despite these advantages, FLC-based approaches face several limitations. The design of fuzzy rules and membership functions is often heuristic and requires expert knowledge, limiting scalability and reproducibility. Moreover, stability guarantees are difficult to establish, and response times may be insufficient for rapidly changing cardiovascular conditions. While DRL-enhanced FLC systems have demonstrated improved tracking performance compared to PID controllers, they introduce additional challenges, including high computational complexity, data dependency, and limited generalization to unseen patient conditions.

These limitations highlight the need for more systematic and theoretically grounded control strategies, motivating the development of robust and adaptive control methods discussed in the following subsections.

\subsection{Robust Control}
\label{Robust}

Robust control strategies are designed to maintain stable and safe LVAD operation in the presence of model uncertainty, external disturbances, and patient-specific variability. Unlike classical controllers, which rely on fixed parameters, robust control explicitly accounts for bounded uncertainties and aims to ensure acceptable performance under worst-case conditions. In LVAD applications, this involves satisfying critical physiological constraints, such as maintaining cardiac output, regulating flow and pressure, preventing ventricular suction, and ensuring energy-efficient pump operation.

Optimal control methods constitute an important class of robust strategies, where a predefined cost function, typically balancing tracking performance, control effort, and physiological safety, is minimized subject to system constraints. For example, PI-based optimal controllers have been used to dynamically regulate motor current in axial and rotary pumps \cite{giridharan2003control}, while state-feedback controllers tuned via particle swarm optimization have demonstrated improved hemodynamic performance \cite{yazdi2013improvement}. In addition, $H_{\infty}$ control techniques have been employed to achieve robust regulation under parameter uncertainty and disturbances by minimizing the worst-case gain from disturbance to output \cite{bakouri2022optimal}.

Despite their theoretical guarantees, optimal and $H_{\infty}$-based controllers depend strongly on the availability of accurate cardiovascular models and well-defined uncertainty bounds. In practice, these requirements are difficult to satisfy due to the complex, nonlinear, and patient-specific nature of cardiovascular dynamics, limiting their clinical applicability.

To overcome these limitations, sliding mode control (SMC) has been proposed as a nonlinear robust control technique capable of maintaining performance under significant uncertainties. SMC-based LVAD controllers have been shown to mimic physiological mechanisms such as the Frank–Starling response and restore abnormal hemodynamic conditions in heart failure simulations \cite{bakouri2014sliding}. Further developments, including pole-placement-based SMC, have improved perfusion regulation and system responsiveness \cite{bakouri2015nonlinear}.

However, SMC introduces its own challenges, most notably chattering effects caused by high-frequency switching, which may lead to excessive control effort and potential mechanical wear in LVAD systems. Additionally, standard SMC formulations lack adaptability to long-term physiological changes, limiting their effectiveness in dynamic clinical environments. These limitations highlight the need for control strategies that can both handle uncertainty and adapt to time-varying patient conditions, motivating the development of adaptive control approaches discussed in the following subsection.

\subsection{Adaptive and Intelligent Control}
\label{Adaptive}

To address the limitations of both classical and robust control strategies in handling time-varying cardiovascular dynamics, \textbf{adaptive control methods} have been proposed for LVAD systems. Unlike fixed-gain controllers, adaptive approaches continuously update control parameters in real time to accommodate changes in patient condition, physiological demand, and system dynamics, without requiring an exact mathematical model.

In LVAD applications, model-free adaptive controllers have demonstrated the ability to regulate pump speed in response to varying physiological conditions \cite{son2022model, chang2011model}. Feedback-based adaptive strategies utilizing measurable signals such as motor current or power consumption have also been explored to estimate hemodynamic states and adjust pump operation accordingly \cite{wang2012feedback, son2020modelling}. Furthermore, adaptive control frameworks inspired by the Frank–Starling mechanism have been developed to maintain physiologically consistent flow regulation under dynamic conditions \cite{gaddum2014starling}.

Despite their ability to handle time-varying dynamics, adaptive controllers often exhibit slow convergence and limited responsiveness to rapid physiological changes, which may compromise performance in critical scenarios such as sudden activity shifts or pathological events.

To overcome these limitations, \textbf{intelligent control techniques} have been introduced, leveraging data-driven models to capture complex nonlinear relationships within the cardiovascular system. Artificial neural networks (ANNs), as universal function approximators, have been widely used for modeling and control of LVAD systems \cite{hornik1989multilayer, kim1997control, fetanat2021sensorless, fetanat2021fully, bahl2023explainable, just2024artificial}. These approaches enable the estimation of unmeasurable physiological variables and support advanced control strategies, including applications in biventricular support systems to balance systemic and pulmonary circulation \cite{ng2017application, ng2018application}.

More recently, deep reinforcement learning (DRL) has emerged as a powerful framework for optimizing LVAD control policies through interaction with the system environment. DRL-based controllers have demonstrated improved adaptability and performance under diverse physiological conditions \cite{wang2024aortic, berger2025enhancing, al2021robustness}. However, these methods require large amounts of high-quality training data, are computationally intensive, and often lack interpretability, posing significant challenges for real-time clinical deployment.

These limitations motivate the development of more structured AI/ML-based control frameworks, which are discussed in the following subsection.

\subsection{AI and Machine Learning-Based Approaches}
\label{AI}

Despite significant advances in classical modeling and control, conventional LVAD systems remain limited in their ability to adapt to dynamic and patient-specific physiological conditions. This limitation can lead to adverse outcomes such as ventricular suction, insufficient perfusion, pulmonary congestion, hemolysis, and thrombus formation \cite{belkacem2023optimizing, vollkron2004development, gregory2024mechanical, reesink2007suction}. These challenges highlight the need for data-driven approaches capable of capturing complex, nonlinear, and patient-specific cardiovascular dynamics.

Machine learning (ML) techniques have emerged as a promising solution by enabling the extraction of patterns and relationships directly from physiological data, without requiring explicit mathematical models. In the context of LVAD systems, ML applications can be broadly categorized into three functional domains: (i) physiological state detection, (ii) risk prediction and clinical decision support, and (iii) control and optimization.

\subsubsection*{A. Physiological State Detection and Monitoring}

One of the earliest and most critical applications of ML in LVAD systems is the detection of adverse physiological conditions, particularly ventricular suction. Early approaches employed classification and regression trees to identify pump states with high sensitivity and specificity \cite{karantonis2006identification}. Subsequent methods utilized signal processing and statistical learning techniques, such as discriminant analysis, to detect time- and frequency-domain variations in pump flow signals \cite{ferreira2006discriminant}. 

\begin{figure*}[t!]
\centering
\includegraphics[width=0.6\textwidth]{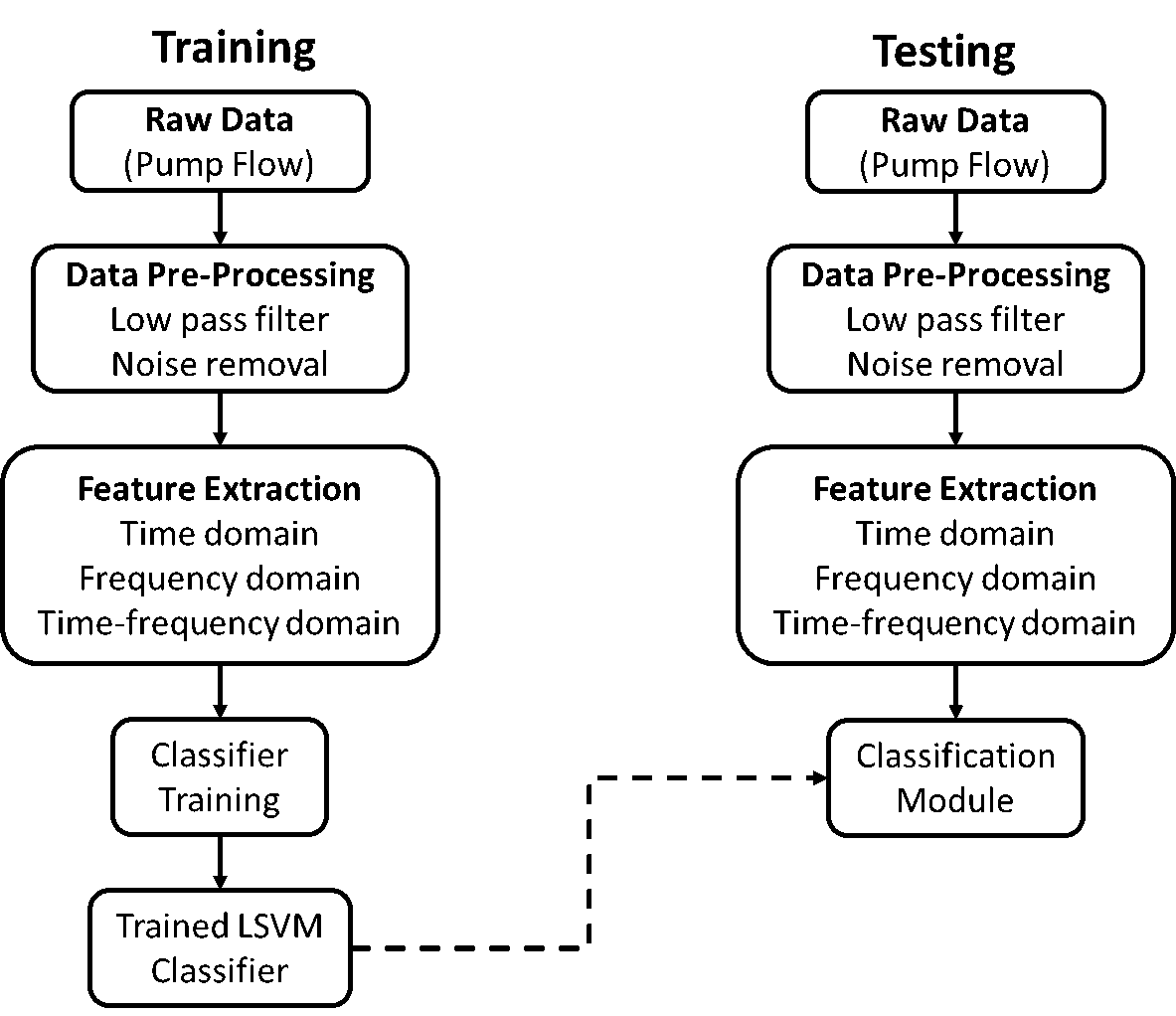}
\caption{Flowchart of LSVM-based suction detection algorithm \cite{ayala2012tuning}.}
\label{Ali_f1}
\end{figure*}

Neural network-based classifiers further improved detection performance by learning complex nonlinear relationships from patient data, achieving sensitivity and specificity exceeding 98\% in suction detection tasks \cite{karantonis2008noninvasive}. More recently, support vector machine-based approaches, including Lagrangian SVMs, have been proposed to enhance classification robustness through structured preprocessing and feature extraction pipelines, as illustrated in Fig.~\ref{Ali_f1} \cite{wang2013suction}.

While these methods significantly improve detection accuracy, they remain dependent on feature engineering and may struggle with generalization across diverse patient populations.

\subsubsection*{B. Risk Prediction and Clinical Decision Support}

Beyond real-time monitoring, ML has been increasingly applied to predict long-term clinical outcomes and support therapeutic decision-making. Predictive models have been developed to assess myocardial recovery, identify candidates for LVAD explantation, and optimize transplant strategies \cite{wilcox2020heart, mann2012myocardial}. 

Recent studies have demonstrated the superiority of ML-based approaches over traditional statistical models. For instance, ensemble and deep learning models have achieved higher predictive accuracy in identifying patient subgroups likely to benefit from therapies such as cardiac resynchronization \cite{cikes2019machine, howell2021using, de2023machine}. Similarly, comparative analyses of multiple ML algorithms have shown improved performance (AUC up to 0.824) in predicting recovery outcomes in LVAD-supported patients \cite{topkara2022machine}.

In risk assessment, large-scale ML models such as MARKER-HF have leveraged clinical datasets to predict post-implantation mortality with improved accuracy compared to conventional scoring systems like INTERMACS \cite{park2024machine}. However, these approaches are often limited by data heterogeneity, missing clinical variables, and lack of interpretability, which hinder their widespread adoption in clinical practice.

\subsubsection*{C. Data-Driven Control and Optimization}

ML techniques have also been explored for enhancing LVAD control strategies. Early efforts focused on using AI methods for controller parameter tuning, improving performance over manually tuned systems \cite{aly2011pid, ayala2012tuning}. More advanced approaches, such as iterative learning control (ILC), have been proposed to optimize pump speed trajectories across cardiac cycles by learning from previous system behavior \cite{ketelhut2017iterative}.

\begin{figure*}[t!]
\centering
\includegraphics[width=0.8\textwidth]{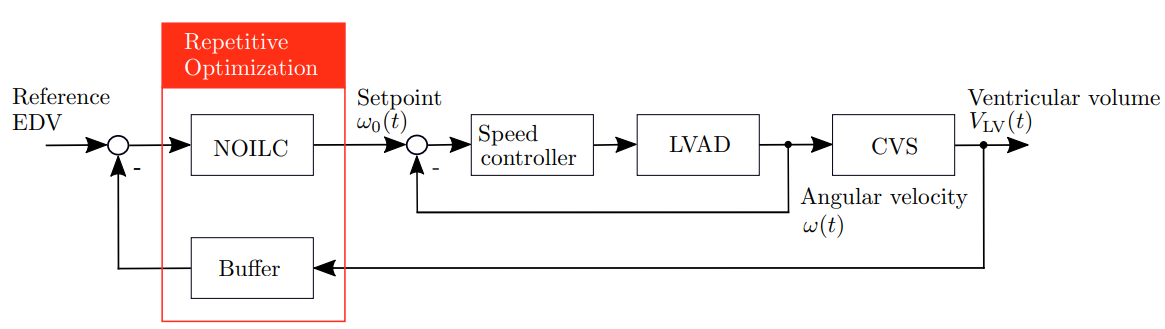}
\caption{Iterative learning control framework for LVAD speed regulation \cite{ketelhut2017iterative}.}
\label{Ali_f2}
\end{figure*}

Variants of ILC incorporate nonlinear system dynamics, parallel feedback structures, and adaptability to varying cardiac cycle durations \cite{ruschen2017minimizing, ketelhut2018iterative, ketelhut2019iterative}. Model-free extensions, such as physiologic data-driven ILC (PDD-ILC), further enhance adaptability by directly following treatment-driven flow trajectories \cite{magkoutas2022physiologic}.

Hybrid intelligent controllers combining ML with fuzzy logic or evolutionary optimization have also demonstrated improved performance. For example, supervised adaptive fuzzy controllers achieve faster response times and improved suction margins compared to conventional methods \cite{azizkhani2022supervised}, while genetic algorithms have been used to optimize control parameters based on treatment-specific objectives \cite{magkoutas2023genetic}.

Despite these advancements, ML-based control approaches face significant challenges, including high data requirements, computational complexity, limited interpretability, and lack of robustness to unseen physiological conditions. These limitations currently restrict their translation into fully autonomous clinical LVAD systems.

\medskip
Overall, while AI and ML techniques offer substantial potential to overcome the limitations of traditional modeling and control approaches, a critical gap remains between algorithmic performance and clinical applicability. Bridging this gap requires the development of hybrid frameworks that integrate physiological knowledge with data-driven learning, ensuring both robustness and interpretability in real-world LVAD applications.

\begin{table*}[!t]
\centering
\caption{Comparative analysis of LVAD control strategies in terms of adaptability, robustness, and clinical applicability.}
\label{tab:control_strategies}

\setlength{\tabcolsep}{3pt}
\renewcommand{\arraystretch}{1.2}

\begin{tabular}{|p{2cm}|p{2cm}|p{2.1cm}|p{2cm}|p{2cm}|p{2cm}|p{2cm}|}
\hline
\textbf{Control Strategy} & \textbf{Model Dependency} & \textbf{Adaptability} & \textbf{Sensor Requirement} & \textbf{Robustness} & \textbf{Real-Time Feasibility} & \textbf{Key Limitation / Clinical Barrier} \\
\hline

\textbf{PID Control} &
High &
Low &
Requires pressure/flow sensors &
Low--Moderate &
High &
Poor performance under nonlinear and time-varying conditions; reliance on invasive sensors \\
\hline

\textbf{Fuzzy Logic Control (FLC)} &
Low (model-free) &
Moderate &
Can be sensorless or minimal sensors &
Moderate &
High &
Rule tuning is heuristic; lacks stability guarantees in critical conditions \\
\hline

\textbf{Robust Control (H$_\infty$, SMC)} &
High &
Low--Moderate &
Requires accurate state estimation &
High &
Moderate &
Sensitive to model inaccuracies; SMC suffers from chattering and high control effort \\
\hline

\textbf{Adaptive Control} &
Moderate &
High &
Often sensor-dependent (flow, power) &
Moderate &
Moderate &
Slow response and stability concerns under rapid physiological changes \\
\hline

\textbf{AI/ML-Based Control} &
Low (data-driven) &
Very High &
Can enable sensorless estimation &
High (if trained well) &
Moderate &
Requires large datasets; limited interpretability and clinical validation \\
\hline

\end{tabular}
\end{table*}

\medskip
Table~\ref{tab:control_strategies} highlights that no single control strategy fully satisfies the requirements of robustness, adaptability, and clinical feasibility in LVAD systems. Classical controllers such as PID offer simplicity and real-time implementation but lack adaptability to nonlinear and time-varying physiological conditions. Advanced methods, including robust and adaptive control, improve stability but remain dependent on accurate models and sensor availability. AI/ML-based approaches provide significant potential for personalization and sensorless control; however, their clinical adoption is limited by data requirements, interpretability concerns, and lack of regulatory validation. These trade-offs underscore the need for hybrid and physiologically informed control strategies that bridge the gap between theoretical performance and clinical applicability.

\section{Future Directions and Challenges}
\label{Future}

Despite significant progress in modeling and control of LVAD systems, several critical challenges continue to hinder their clinical translation. These challenges primarily arise from the complex, nonlinear, and patient-specific nature of cardiovascular dynamics, as well as the limited availability of reliable physiological measurements. Future research must therefore move beyond incremental improvements and focus on fundamentally bridging the gap between theoretical models and real-world clinical applicability.

\subsection{Bridging the Model-Reality Gap}

A central challenge in LVAD research is the discrepancy between mathematical models and in vivo physiological behavior. Most existing models rely on simplifying assumptions that fail to capture patient-specific variability, time-varying dynamics, and complex pump–heart interactions.

\textbf{Koopman operator theory} has recently emerged as a promising approach for addressing this limitation. By lifting nonlinear system dynamics into a higher-dimensional linear space, Koopman-based methods enable the application of linear analysis and control techniques to inherently nonlinear cardiovascular systems. This framework offers two key advantages: (i) the ability to model complex fluid–structure–electromechanical interactions in a unified manner, and (ii) the development of reduced-order models that retain essential system dynamics while remaining computationally efficient. However, practical implementation remains challenging due to the need for high-quality data and appropriate observable selection.

\subsection{Patient-Specific Modeling via System Identification}

Another major limitation of current LVAD models is their lack of personalization. Inter-patient variability in cardiovascular physiology significantly affects device performance, making generic models insufficient for reliable control.

System identification techniques offer a pathway toward patient-specific modeling by constructing models directly from measured data. Subspace identification methods enable real-time extraction of state-space representations from sensor data, facilitating the design of individualized controllers. Similarly, Hammerstein–Wiener models provide a structured approach for capturing nonlinear relationships between physiological inputs and outputs, improving model fidelity across diverse operating conditions.

Despite their potential, these approaches depend heavily on the availability of accurate and continuous measurements, which remains a major limitation due to the lack of reliable implantable sensors.

\subsection{Uncertainty-Aware and Stochastic Control}

A critical gap in existing LVAD control strategies is the limited consideration of uncertainty arising from physiological variability, patient behavior, and environmental factors. Deterministic control approaches are inherently insufficient to address these uncertainties, particularly in long-term, real-world applications.

Stochastic control frameworks provide a more realistic representation by explicitly modeling uncertainty through probabilistic methods. Potential approaches include stochastic state-space models, Markov decision processes, and Bayesian networks to capture dependencies among physiological variables. These methods can enable more robust and adaptive control strategies capable of handling unpredictable variations in patient state.

However, the implementation of stochastic control in LVAD systems faces significant challenges, including the need for large-scale longitudinal datasets, computational complexity, and the difficulty of accurately characterizing probability distributions in clinical settings.

\subsection{Next-Generation AI and Hybrid Modeling Frameworks}

While current AI-based approaches have demonstrated promising results, they remain limited by data dependency, lack of interpretability, and poor generalization. Future research must therefore focus on integrating data-driven methods with physiological knowledge.

Physics-informed neural networks (PINNs) represent a promising direction by embedding governing physical laws into learning frameworks, enabling improved generalization and reduced data requirements. Similarly, advanced deep learning architectures such as Long Short-Term Memory (LSTM) networks and convolutional neural networks (CNNs) can be leveraged to model temporal and spatial patterns in physiological data.

A particularly promising direction lies in \textbf{hybrid modeling frameworks}, which combine mechanistic models with data-driven learning. Such approaches can exploit the strengths of both paradigms—interpretability from physics-based models and adaptability from machine learning—potentially leading to clinically viable LVAD control systems.

\subsection{Towards Clinically Deployable LVAD Control Systems}

Ultimately, the translation of advanced modeling and control strategies into clinical practice requires addressing several practical constraints, including sensor limitations, computational efficiency, safety certification, and real-time implementation.

Future LVAD systems must therefore be designed with a focus on:
\begin{itemize}
    \item Sensorless or minimally invasive estimation techniques
    \item Real-time capable control algorithms
    \item Robustness to patient variability and uncertainty
    \item Clinical interpretability and safety assurance
\end{itemize}

Bridging these gaps will require interdisciplinary collaboration across biomedical engineering, control theory, data science, and clinical practice. Only through such integration can next-generation LVAD systems achieve truly autonomous, safe, and patient-specific operation.

\section{Conclusions}
\label{Conclusion}

This review has provided a comprehensive and critical synthesis of modeling and control strategies for left ventricular assist devices (LVADs), with a focus on their interaction with the cardiovascular system (CVS). The analysis demonstrates that while deterministic modeling approaches such as time-varying elastance and Windkessel-based lumped parameter models offer valuable physiological insight and support controller design, they remain limited in their ability to capture patient-specific variability and long-term dynamics. Emerging stochastic and data-driven modeling techniques partially address these limitations but introduce challenges related to data availability, interpretability, and computational complexity.

From a control perspective, classical strategies such as PID controllers provide simplicity and ease of implementation but lack adaptability to nonlinear and time-varying physiological conditions. Advanced methods, including fuzzy logic, robust control, and adaptive control, improve performance under uncertainty but often involve trade-offs between stability, responsiveness, and implementation complexity. More recently, artificial intelligence and machine learning-based approaches have demonstrated strong potential in capturing nonlinear dynamics, enabling predictive capabilities, and supporting personalized therapy. However, their clinical applicability remains constrained by data requirements, limited interpretability, and challenges in real-time deployment.

A key insight from this review is that no single modeling or control paradigm fully satisfies the competing requirements of physiological accuracy, robustness, adaptability, and clinical feasibility. This highlights the need for hybrid frameworks that integrate mechanistic understanding with data-driven learning. Such approaches offer a promising pathway toward the development of next-generation LVAD systems capable of autonomous, patient-specific, and physiologically responsive operation.

Ultimately, advancing LVAD technology requires bridging the gap between theoretical developments and clinical implementation. Future progress will depend on the integration of robust modeling, adaptive control, reliable sensing (or sensorless estimation), and interpretable AI, supported by interdisciplinary collaboration between engineers, clinicians, and data scientists. Addressing these challenges is essential for improving patient outcomes and realizing the full potential of LVAD therapy in the management of advanced heart failure.

\backmatter

\textbf{Conflict of Interest Statement}

The authors declare that there is no conflict of interest associated with this article.

\bibliographystyle{sn_mathphys_num} 
\bibliography{sn_bibliography} 

@article{shim2004mathematical,
  title={Mathematical modeling of cardiovascular system dynamics using a lumped parameter method},
  author={Shim, Eun Bo and Sah, Jong Youb and Youn, Chan Hyun},
  journal={The Japanese journal of physiology},
  volume={54},
  number={6},
  pages={545--553},
  year={2004},
  publisher={THE PHYSIOLOGICAL SOCIETY OF JAPAN}
}

@article{grzyb2024artificial,
  title={Artificial intelligence approaches for predicting the risks of durable mechanical circulatory support therapy and cardiac transplantation},
  author={Grzyb, Chloe and Du, Dongping and Nair, Nandini},
  journal={Journal of Clinical Medicine},
  volume={13},
  number={7},
  pages={2076},
  year={2024},
  publisher={MDPI}
}

@article{balcioglu2024role,
  title={The Role of Artificial Intelligence and Machine Learning in the Prediction of Right Heart Failure after Left Ventricular Assist Device Implantation: A Comprehensive Review},
  author={Balcioglu, Ozlem and Ozgocmen, Cemre and Ozsahin, Dilber Uzun and Yagdi, Tahir},
  journal={Diagnostics},
  volume={14},
  number={4},
  pages={380},
  year={2024},
  publisher={MDPI}
}

@article{suero2023optimization,
  title={Optimization of Left Ventricular Assist Device Support},
  author={Suero, Arantxa G and Xie, Lola X},
  journal={Texas Heart Institute Journal},
  volume={50},
  number={4},
  pages={e238231},
  year={2023},
  publisher={Texas Heart{\textregistered} Institute, Houston}
}

@article{alomari2012developments,
  title={Developments in control systems for rotary left ventricular assist devices for heart failure patients: a review},
  author={AlOmari, Abdul-Hakeem H and Savkin, Andrey V and Stevens, Michael and Mason, David G and Timms, Daniel L and Salamonsen, Robert F and Lovell, Nigel H},
  journal={Physiological measurement},
  volume={34},
  number={1},
  pages={R1},
  year={2012},
  publisher={IOP Publishing}
}

@article{vseman2024computational,
  title={Computational modelling of valvular heart disease: haemodynamic insights and clinical implications},
  author={{\v{S}}eman, Michael and Stephens, Andrew F and Kaye, David M and Gregory, Shaun D and Stub, Dion},
  journal={Frontiers in Bioengineering and Biotechnology},
  volume={12},
  pages={1462542},
  year={2024},
  publisher={Frontiers Media SA}
}

@article{capoccia2015development,
  title={Development and characterization of the arterial W indkessel and its role during left ventricular assist device assistance},
  author={Capoccia, Massimo},
  journal={Artificial organs},
  volume={39},
  number={8},
  pages={E138--E153},
  year={2015},
  publisher={Wiley Online Library}
}

@article{fernandes2023modeling,
  title={Modeling the Five-Element Windkessel Model with Simultaneous Utilization of Blood Viscoelastic Properties for FFR Achievement: A Proof-of-Concept Study},
  author={Fernandes, Maria and Sousa, Luisa C and Ant{\'o}nio, Carlos A Concei{\c{c}}{\~a}o and Pinto, S{\'o}nia IS},
  journal={Mathematics},
  volume={11},
  number={24},
  pages={4877},
  year={2023},
  publisher={MDPI}
}

@article{ujiie2022solitonic,
  title={Solitonic Windkessel model for intracranial aneurysm},
  author={Ujiie, Hiroshi and Iwata, Yoritaka},
  journal={Brain Sciences},
  volume={12},
  number={8},
  pages={1016},
  year={2022},
  publisher={MDPI}
}

@book{noordergraaf2012circulatory,
  title={Circulatory system dynamics},
  author={Noordergraaf, Abraham},
  volume={1},
  year={2012},
  publisher={Elsevier}
}

@article{abdi2015lumped,
  title={A lumped parameter mathematical model to analyze the effects of tachycardia and bradycardia on the cardiovascular system},
  author={Abdi, Mohsen and Karimi, Alireza and Navidbakhsh, Mahdi and Pirzad Jahromi, Gila and Hassani, Kamran},
  journal={International Journal of Numerical Modelling: Electronic Networks, Devices and Fields},
  volume={28},
  number={3},
  pages={346--357},
  year={2015},
  publisher={Wiley Online Library}
}

@book{formaggia2010cardiovascular,
  title={Cardiovascular Mathematics: Modeling and simulation of the circulatory system},
  author={Formaggia, Luca and Quarteroni, Alfio and Veneziani, Allesandro},
  volume={1},
  year={2010},
  publisher={Springer Science \& Business Media}
}

@article{fu2015influence,
  title={The influence of hemodynamics on the ulceration plaques of carotid artery stenosis},
  author={Fu, Yulin and Qiao, Aike and Jin, Long},
  journal={Journal of Mechanics in Medicine and Biology},
  volume={15},
  number={01},
  pages={1550008},
  year={2015},
  publisher={World Scientific}
}

@inproceedings{bahloul2023fractional,
  title={Fractional-order Modified Windkessel Model of the Human Arterial Vascular System},
  author={Bahloul, Mohamed A and Aboelkassem, Yasser and Belkhatir, Zehor and Laleg-Kirati, Taous-Meriem},
  booktitle={2023 IEEE Biomedical Circuits and Systems Conference (BioCAS)},
  pages={1--5},
  year={2023},
  organization={IEEE}
}

@inproceedings{aka2022complete,
  title={A complete LPM-CFD Coupling on a Dummy Aortic Model},
  author={Aka, Ibrahim Basar and Yildirim, Canberk},
  booktitle={2022 Medical Technologies Congress (TIPTEKNO)},
  pages={1--4},
  year={2022},
  organization={IEEE}
}

@article{garven2025pediatric,
  title={Pediatric Cardiovascular Multiscale Modeling using a Functional Mock-up Interface},
  author={Garven, Ellen E and Kung, Ethan and Stevens, Randy M and Throckmorton, Amy L},
  journal={Cardiovascular engineering and technology},
  pages={1--9},
  year={2025},
  publisher={Springer}
}

@article{liu2024development,
  title={Development of a Lumped Parameter Model of Human Whole Body Circulatory Loop},
  author={Liu, Xun and Mo, Chenghuai and Li, Jiaxing and Yu, Hongyi and Hu, Sheng and Zhu, Da and Zhang, Puming and Li, Yonghua},
  journal={IEEE Access},
  year={2024},
  publisher={IEEE}
}

@inproceedings{rubtsova2021lumped,
  title={Lumped parameter model of the cardiovascular system with baroreflex},
  author={Rubtsova, E and Telyshev, D},
  booktitle={Journal of Physics: Conference Series},
  volume={2091},
  number={1},
  pages={012024},
  year={2021},
  organization={IOP Publishing}
}

@article{son2020modelling,
  title={Modelling and control of a failing heart managed by a left ventricular assist device},
  author={Son, Jeongeun and Du, Dongping and Du, Yuncheng},
  journal={Biocybernetics and biomedical engineering},
  volume={40},
  number={1},
  pages={559--573},
  year={2020},
  publisher={Elsevier}
}

@inproceedings{ferreira2005nonlinear,
  title={A nonlinear state-space model of a combined cardiovascular system and a rotary pump},
  author={Ferreira, Antonio and Chen, Shaohui and Simaan, Marwan A and Boston, J Robert and Antaki, James F},
  booktitle={Proceedings of the 44th IEEE Conference on Decision and Control},
  pages={897--902},
  year={2005},
  organization={IEEE}
}

@inproceedings{ferreira2005dynamical,
  title={A Dynamical State Space Representation and Performance Analysis of a Feedback Controlled Rotary Left Ventricular Assist Device},
  author={Ferreira, Antonio and Chen, Shaohui and Galati, David G and Simaan, Marwan A and Antaki, James F},
  booktitle={ASME International Mechanical Engineering Congress and Exposition},
  volume={42169},
  pages={617--626},
  year={2005}
}

@article{schima1990computer,
  title={Computer simulation of the circulatory system during support with a rotary blood pump},
  author={SCHIMA, HEINRICH and HONIGSCHNABH, JOHANN and WOHGAN G, TRUBH and THOMA, HERWIG and others},
  journal={ASAIO Journal},
  volume={36},
  number={3},
  pages={M252--253},
  year={1990},
  publisher={LWW}
}

@article{westerhof1969analog,
  title={Analog studies of the human systemic arterial tree},
  author={Westerhof, Nicolaas and Bosman, Frederik and De Vries, Cornelis J and Noordergraaf, Abraham},
  journal={Journal of biomechanics},
  volume={2},
  number={2},
  pages={121--143},
  year={1969},
  publisher={Elsevier}
}

@article{luo2007using,
  title={Using a human cardiopulmonary model to study and predict normal and diseased ventricular mechanics, septal interaction, and atrio-ventricular blood flow patterns},
  author={Luo, C and Ware, DL and Zwischenberger, JB and Clark, JW},
  journal={Cardiovascular engineering},
  volume={7},
  pages={17--31},
  year={2007},
  publisher={Springer}
}

@article{ferrari2005development,
  title={Development of hybrid (numerical-physical) models of the cardiovascular system: Numerical-electrical and numerical hydraulic applications},
  author={Ferrari, GIANFRANCO and Kozarski, MACIEJ and De Lazzari, CLAUDIO and G{\'o}rczy{\'n}ska, K and Darowski, MAREK and Tosti, GIANCARLO},
  journal={Biocybernetics and Biomedical Engineering},
  volume={25},
  number={4},
  pages={3--15},
  year={2005}
}

@article{ferrari2009role,
  title={Role and applications of circulatory models in cardiovascular pathophysiology},
  author={Ferrari, Gianfranco and Kozarski, Maciej and De Lazzari, Claudio and G{\'o}rczy{\'n}ska, Krystyna and Pa{\l}ko, Krzysztof J and Zieli{\'n}ski, Krzysztof and Di Molfetta, Arianna and Darowski, Marek},
  journal={Biocybern Biomed Eng},
  volume={29},
  pages={3--24},
  year={2009}
}

@article{bahloul2019fractional,
  title={Fractional order models of arterial windkessel as an alternative in the analysis of the left ventricular afterload},
  author={Bahloul, Mohamed A and Kirati, Taous-Meriem Laleg},
  journal={arXiv preprint arXiv:1908.05239},
  year={2019}
}

@article{burattini2007development,
  title={Development of systemic arterial mechanical properties from infancy to adulthood interpreted by four-element windkessel models},
  author={Burattini, Roberto and Di Salvia, Paola Oriana},
  journal={Journal of Applied Physiology},
  volume={103},
  number={1},
  pages={66--79},
  year={2007},
  publisher={American Physiological Society}
}

@article{burattini1998complex,
  title={Complex and frequency-dependent compliance of viscoelastic windkessel resolves contradictions in elastic windkessels},
  author={Burattini, Roberto and Natalucci, Silvia},
  journal={Medical engineering \& physics},
  volume={20},
  number={7},
  pages={502--514},
  year={1998},
  publisher={Elsevier}
}

@article{grant1987characterization,
  title={Characterization of pulmonary arterial input impedance with lumped parameter models},
  author={Grant, BJ and Paradowski, LINDA J},
  journal={American Journal of Physiology-Heart and Circulatory Physiology},
  volume={252},
  number={3},
  pages={H585--H593},
  year={1987}
}

@article{stergiopulos1999total,
  title={Total arterial inertance as the fourth element of the windkessel model},
  author={Stergiopulos, Nikos and Westerhof, Berend E and Westerhof, Nico},
  journal={American Journal of Physiology-Heart and Circulatory Physiology},
  volume={276},
  number={1},
  pages={H81--H88},
  year={1999},
  publisher={American Physiological Society Bethesda, MD}
}

@article{frank1899erste,
  title={Erste Abhandlung. Mathematische Analyse},
  author={Frank, O and des arteriellen Pulses, Die Grundform},
  journal={Zeitschrift fur Biologie},
  volume={37},
  pages={485--526},
  year={1899}
}

@article{westerhof1971artificial,
  title={An artificial arterial system for pumping hearts.},
  author={Westerhof, Nicolaas and Elzinga, GIJS and Sipkema, Pieter},
  journal={Journal of applied physiology},
  volume={31},
  number={5},
  pages={776--781},
  year={1971}
}

@article{kokalari2013review,
  title={Review on lumped parameter method for modeling the blood flow in systemic arteries},
  author={Kokalari, Isidor and Karaja, Theodhor and Guerrisi, Maria},
  journal={Journal of Biomedical Science and Engineering},
  year={2013},
  publisher={Scientific Research Publishing}
}

@inproceedings{tan2016modelling,
  title={Modelling of human cardiovascular system in ventricular assist device simulation},
  author={Tan, Kean Eng and Yahya, Samer and Almurib, Haider AF and Moghavvemi, Mahmoud},
  booktitle={2016 IEEE Industrial Electronics and Applications Conference (IEACon)},
  pages={304--311},
  year={2016},
  organization={IEEE}
}

@article{stergiopulos1996determinants,
  title={Determinants of stroke volume and systolic and diastolic aortic pressure},
  author={Stergiopulos, NIKOS and Meister, JEAN-JACQUES and Westerhof, NICO},
  journal={American Journal of Physiology-Heart and Circulatory Physiology},
  volume={270},
  number={6},
  pages={H2050--H2059},
  year={1996}
}

@article{vasudevan2022application,
  title={Application of mathematical modeling to quantify ventricular contribution following durable left ventricular assist device support},
  author={Vasudevan, Viswajith S and Rajagopal, Keshava and Antaki, James F},
  journal={Applications in Engineering Science},
  volume={11},
  pages={100107},
  year={2022},
  publisher={Elsevier}
}

@article{hedayati2024elastance,
  title={Elastance-based nonlinear dynamical controller synthesis for the in-vitro cardiovascular circulatory system: A backstepping approach},
  author={Hedayati, Shahrzad and Abbasi Nozari, Hasan and Sadati Rostami, Seyed Jalil and Castaldi, Paolo and Noshad, Erfan},
  journal={Journal of Vibration and Control},
  pages={10775463241305679},
  year={2024},
  publisher={SAGE Publications Sage UK: London, England}
}

@article{ochsner2017novel,
  title={A novel mean-value model of the cardiovascular system including a left ventricular assist device},
  author={Ochsner, Gregor and Amacher, Raffael and Schmid Daners, Marianne},
  journal={Cardiovascular engineering and technology},
  volume={8},
  pages={120--130},
  year={2017},
  publisher={Springer}
}

@inproceedings{ferreira2007rule,
  title={A rule-based controller based on suction detection for rotary blood pumps},
  author={Ferreira, Antonio and Boston, J Robert and Antaki, James F},
  booktitle={2007 29th Annual International Conference of the IEEE Engineering in Medicine and Biology Society},
  pages={3978--3981},
  year={2007},
  organization={IEEE}
}

@article{segers2003systemic,
  title={Systemic and pulmonary hemodynamics assessed with a lumped-parameter heart-arterial interaction model},
  author={Segers, Patrick and Stergiopulos, Nikos and Westerhof, Nico and Wouters, Patrick and Kolh, Philippe and Verdonck, Pascal},
  journal={Journal of engineering mathematics},
  volume={47},
  pages={185--199},
  year={2003},
  publisher={Springer}
}

@inproceedings{simaan2008modeling,
  title={Modeling and control of the heart left ventricle supported with a rotary assist device},
  author={Simaan, Marwan A},
  booktitle={2008 47th IEEE Conference on Decision and Control},
  pages={2656--2661},
  year={2008},
  organization={IEEE}
}

@article{son2019stochastic,
  title={Stochastic modeling and dynamic analysis of the cardiovascular system with rotary left ventricular assist devices},
  author={Son, Jeongeun and Du, Dongping and Du, Yuncheng},
  journal={Mathematical Problems in Engineering},
  volume={2019},
  number={1},
  pages={7179317},
  year={2019},
  publisher={Wiley Online Library}
}

@inproceedings{tan2024dynamic,
  title={A Dynamic Model of Multi-state LVAD Based on LSTM Neural Network},
  author={Tan, Aiping and Mu, Ying and Yu, Wenqian and Liang, Chenxi and Chen, Yanfeng},
  booktitle={International Conference on Intelligent Computing},
  pages={203--214},
  year={2024},
  organization={Springer}
}

@article{shi2006numerical,
  title={Numerical simulation of cardiovascular dynamics with left heart failure and in-series pulsatile ventricular assist device},
  author={Shi, Yubing and Korakianitis, Theodosios},
  journal={Artificial organs},
  volume={30},
  number={12},
  pages={929--948},
  year={2006},
  publisher={Wiley Online Library}
}

@article{sarnoff1954ventricular,
  title={Ventricular function: I. Starling's law of the heart studied by means of simultaneous right and left ventricular function curves in the dog},
  author={Sarnoff, Stanley J and Berglund, Erik},
  journal={Circulation},
  volume={9},
  number={5},
  pages={706--718},
  year={1954},
  publisher={Am Heart Assoc}
}

@article{suga1974instantaneous,
  title={Instantaneous pressure-volume relationships and their ratio in the excised, supported canine left ventricle},
  author={Suga, Hiroyuki and Sagawa, Kiichi},
  journal={Circulation research},
  volume={35},
  number={1},
  pages={117--126},
  year={1974},
  publisher={Am Heart Assoc}
}

@article{yin2024real,
  title={Real-time regurgitation estimation in percutaneous left ventricular assist device fully supported condition using an unscented Kalman filter},
  author={Yin, Anyun and Wen, Biyang and Xie, Qilian and Dai, Ming},
  journal={Physiological Measurement},
  year={2024}
}

@article{tsao2022heart,
  title={Heart disease and stroke statistics—2022 update: a report from the American Heart Association},
  author={Tsao, Connie W and Aday, Aaron W and Almarzooq, Zaid I and Alonso, Alvaro and Beaton, Andrea Z and Bittencourt, Marcio S and Boehme, Amelia K and Buxton, Alfred E and Carson, April P and Commodore-Mensah, Yvonne and others},
  journal={Circulation},
  volume={145},
  number={8},
  pages={e153--e639},
  year={2022},
  publisher={Am Heart Assoc}
}

@article{carpenter2013brief,
  title={A brief review of ventricular assist devices and a recommended protocol for pathology evaluations},
  author={Carpenter, Brian A and Gonzalez, Christian J and Jessen, Staci L and Moore, Erica J and Thrapp, Amber N and Weeks, Brad R and Clubb Jr, Fred J},
  journal={Cardiovascular Pathology},
  volume={22},
  number={5},
  pages={408--415},
  year={2013},
  publisher={Elsevier}
}

@article{slaughter2009advanced,
  title={Advanced heart failure treated with continuous-flow left ventricular assist device},
  author={Slaughter, Mark S and Rogers, Joseph G and Milano, Carmelo A and Russell, Stuart D and Conte, John V and Feldman, David and Sun, Benjamin and Tatooles, Antone J and Delgado III, Reynolds M and Long, James W and others},
  journal={New England Journal of Medicine},
  volume={361},
  number={23},
  pages={2241--2251},
  year={2009},
  publisher={Mass Medical Soc}
}

@incollection{simaan2011left,
  title={Left ventricular assist devices: Engineering design considerations},
  author={Simaan, Marwan A and Faragallah, George and Wang, Yu and Divo, Eduardo},
  booktitle={New Aspects of Ventricular Assist Devices},
  volume={2},
  pages={21--42},
  year={2011},
  publisher={Intech Publishers}
}

@article{simaan2008dynamical,
  title={A dynamical state space representation and performance analysis of a feedback-controlled rotary left ventricular assist device},
  author={Simaan, Marwan A and Ferreira, Antonio and Chen, Shaohi and Antaki, James F and Galati, David G},
  journal={IEEE Transactions on Control Systems Technology},
  volume={17},
  number={1},
  pages={15--28},
  year={2008},
  publisher={IEEE}
}

@article{giridharan2002modeling,
  title={Modeling and control of a brushless DC axial flow ventricular assist device},
  author={Giridharan, Guruprasad A and Skliar, Mikhail and Olsen, Donald B and Pantalos, George M},
  journal={ASAIO journal},
  volume={48},
  number={3},
  pages={272--289},
  year={2002},
  publisher={LWW}
}

@article{suga1972mathematical,
  author       = {Suga, H. and Sagawa, K.},
  title        = {Mathematical Interrelationship Between Instantaneous Ventricular Pressure-Volume Ratio and Myocardial Force-Velocity Relation},
  journal      = {Annals of Biomedical Engineering},
  volume       = {1},
  number       = {2},
  pages        = {160--181},
  month        = dec,
  year         = {1972},
  doi          = {10.1007/BF02584205}
}

@article{walley2016left,
  title={Left ventricular function: time-varying elastance and left ventricular aortic coupling.},
  author={Walley, K R},
  journal={Crit Care},
  volume={20},
  number={1},
  pages={270},
  year={2016},
  month={9},
  doi={10.1186/s13054-016-1439-6}
}

@article{fernandez2013object,
  title={Object-oriented modeling and simulation of the closed loop cardiovascular system by using SIMSCAPE},
  author={Fernandez de Canete, J and del Saz-Orozco, P and Moreno-Boza, D and Duran-Venegas, E},
  journal={Comput Biol Med},
  volume={43},
  number={4},
  pages={323--333},
  year={2013},
  month={5},
  doi={10.1016/j.compbiomed.2013.01.007}
}

@inproceedings{wang2022modeling,
  title={Modeling and Design of a Tandem Axial-flow Left Ventricular Assist System},
  author={Wang, L and Du, L and Hu, J},
  booktitle={2022 41st Chinese Control Conference (CCC)},
  year={2022},
  month={7},
  pages={5711--5716},
  organization={IEEE},
  doi={10.23919/CCC55666.2022.9902596}
}

@article{lopez-santana2023computational,
  title={Computational Fluid Dynamics as a Surgical Tool to Optimise the Positioning of the LVAD Outflow Graft for Reducing Aortic Regurgitation},
  author={Lopez-Santana, G B and De Rosis, A and Grant, S W and Venkateswaran, R and Keshmiri, A},
  journal={The Journal of Heart and Lung Transplantation},
  volume={42},
  number={4},
  pages={S416},
  year={2023},
  month={4},
  doi={10.1016/j.healun.2023.02.1077}
}

@article{mcelroy2020impact,
  title={Impact of heart failure severity on ventricular assist device haemodynamics: a computational study},
  author={McElroy, M. and Xenakis, A. and Keshmiri, A.},
  journal={Research on Biomedical Engineering},
  volume={36},
  number={4},
  pages={489--500},
  month={12},
  year={2020},
  doi={10.1007/s42600-020-00088-2}
}

@article{santiago2022design,
  title={Design and execution of a verification, validation, and uncertainty quantification plan for a numerical model of left ventricular flow after LVAD implantation},
  author={Santiago, A. and others},
  journal={PLoS Comput Biol},
  volume={18},
  number={6},
  pages={e1010141},
  month={6},
  year={2022},
  doi={10.1371/journal.pcbi.1010141}
}

@inproceedings{leong2015electromechanics,
  title={Electromechanics modeling of the effects of myocardial infarction on left ventricular function},
  author={Leong, C.-N. and others},
  booktitle={2015 37th Annual International Conference of the IEEE Engineering in Medicine and Biology Society (EMBC)},
  organization={IEEE},
  pages={5684--5687},
  month={8},
  year={2015},
  doi={10.1109/EMBC.2015.7319682}
}

@phdthesis{santiago2018fluid,
  title={Fluid-Electro-Mechanical Model of the Human Heart for Supercomputers},
  author={Santiago, Alfonso},
  year={2018},
  school={Universitat Politècnica de Catalunya}
}

@book{westerhof2019snapshots,
  title={Snapshots of Hemodynamics},
  author={Westerhof, N. and Stergiopulos, N. and Noble, M. I. M. and Westerhof, B. E.},
  edition={Third},
  year={2019},
  publisher={Springer International Publishing},
  doi={10.1007/978-3-319-91932-4}
}

@article{burkhoff2005assessment,
  title={Assessment of systolic and diastolic ventricular properties via pressure-volume analysis: a guide for clinical, translational, and basic researchers},
  author={Burkhoff, D. and Mirsky, I. and Suga, H.},
  journal={American Journal of Physiology-Heart and Circulatory Physiology},
  volume={289},
  number={2},
  pages={H501--H512},
  month={8},
  year={2005},
  doi={10.1152/ajpheart.00138.2005}
}

@article{weber1974determinants,
  title={Determinants of stroke volume in the isolated canine heart},
  author={Weber, K. T. and Janicki, J. S. and Reeves, R. C. and Hefner, L. L. and Reeves, T. J.},
  journal={J Appl Physiol},
  volume={37},
  number={5},
  pages={742--747},
  month={11},
  year={1974},
  doi={10.1152/jappl.1974.37.5.742}
}

@article{sagawa1977end,
  title={End-systolic pressure/volume ratio: A new index of ventricular contractility},
  author={Sagawa, K. and Suga, H. and Shoukas, A. A. and Bakalar, K. M.},
  journal={Am J Cardiol},
  volume={40},
  number={5},
  pages={748--753},
  month={11},
  year={1977},
  doi={10.1016/0002-9149(77)90192-8}
}

@article{suga1971theoretical,
  title={Theoretical analysis of a left-ventricular pumping model based on the systolic time-varying pressure-volume ratio},
  author={Suga, H.},
  journal={IEEE Trans Biomed Eng},
  volume={18},
  number={1},
  pages={47--55},
  month={1},
  year={1971},
  doi={10.1109/tbme.1971.4502789}
}

@article{lankhaar2009modeling,
  title={Modeling the Instantaneous Pressure--Volume Relation of the Left Ventricle: A Comparison of Six Models},
  author={Lankhaar, J.-W. and others},
  journal={Ann Biomed Eng},
  volume={37},
  number={9},
  pages={1710--1726},
  month={9},
  year={2009},
  doi={10.1007/s10439-009-9742-x}
}

@inproceedings{son2019,
  author = {Son, J. and Du, D. and Du, Y.},
  title = {Stochastic Modeling and Control of Circulatory System with a Left Ventricular Assist Device},
  booktitle = {2019 American Control Conference (ACC)},
  publisher = {IEEE},
  month = {7},
  year = {2019},
  pages = {5408--5413},
  doi = {10.23919/ACC.2019.8814745}
}

@article{spanos1998monte,
  title={Monte Carlo Treatment of Random Fields: A Broad Perspective},
  author={Spanos, P. D. and Zeldin, B. A.},
  journal={Applied Mechanics Reviews},
  volume={51},
  number={3},
  pages={219--237},
  month={3},
  year={1998},
  doi={10.1115/1.3098999}
}

@article{petrou2018viscosity,
  title={Viscosity Prediction in a Physiologically Controlled Ventricular Assist Device},
  author={Petrou, A. and Kanakis, M. and Boes, S. and Pergantis, P. and Meboldt, M. and Daners, M. S.},
  journal={IEEE Transactions on Biomedical Engineering},
  volume={65},
  number={10},
  pages={2355--2364},
  month={10},
  year={2018},
  doi={10.1109/TBME.2018.2797424}
}

@article{yoshizawa1994assessing,
  title={Assessing cardiovascular dynamics during ventricular assistance. Use of fuzzy clustering techniques},
  author={Yoshizawa, Makoto and Takeda, Hiroshi and Yambe, Tomoyuki and Nitta, Shin-ichi},
  journal={IEEE Engineering in Medicine and Biology Magazine},
  volume={13},
  number={5},
  pages={687--692},
  year={1994},
  publisher={IEEE}
}

@article{kaufmann1995fuzzy,
  title={Fuzzy control concept for a total artificial heart},
  author={Kaufmann, Ralf and Becker, Kurt and Nix, Christoph and Reul, Helmut and Rau, G{\"u}nter},
  journal={Artificial Organs},
  volume={19},
  number={4},
  pages={355--361},
  year={1995},
  publisher={Wiley Online Library}
}

@inproceedings{santos2019intelligent,
  title={Intelligent Control based on Fuzzy logic embedded in FPGA applied in Ventricular Assist Devices (VADs)},
  author={Santos, Bruno and Le{\~a}o, Tarcisio and Bock, Eduardo},
  booktitle={Proceedings of the 2019 4th International Conference on Robotics, Control and Automation},
  pages={138--143},
  year={2019}
}

@article{kaufmann1997implantable,
  title={The implantable fuzzy controlled Helmholtz-left ventricular assist device: first in vitro testing},
  author={Kaufmann, Ralf and Nix, Christoph and Klein, Martin and Reul, Helmut and Rau, G{\"u}nter},
  journal={Artificial organs},
  volume={21},
  number={2},
  pages={131--137},
  year={1997},
  publisher={Wiley Online Library}
}

@inproceedings{azizkhani2022supervised,
  title={Supervised adaptive fuzzy control of LVAD with pulsatility ratio modulation},
  author={Azizkhani, Milad and Chen, Yue},
  booktitle={2022 IEEE 18th International Conference on Automation Science and Engineering (CASE)},
  pages={2429--2434},
  year={2022},
  organization={IEEE}
}

@article{wang2024aortic,
  title={Aortic Pressure Control Based on Deep Reinforcement Learning for Ex Vivo Heart Perfusion},
  author={Wang, Shangting and Yang, Ming and Liu, Yuan and Yu, Junwen},
  journal={Applied Sciences},
  volume={14},
  number={19},
  pages={8735},
  year={2024},
  publisher={MDPI}
}

@article{li2022intelligent,
  title={Intelligent and strong robust CVS-LVAD control based on soft-actor-critic algorithm},
  author={Li, Te and Cui, Wenbo and Xie, Nan and Li, Heng and Liu, Haibo and Li, Xu and Wang, Yongqing},
  journal={Artificial Intelligence in Medicine},
  volume={128},
  pages={102308},
  year={2022},
  publisher={Elsevier}
}

@article{wang2015rotary,
  title={Rotary blood pump control strategy for preventing left ventricular suction},
  author={Wang, Yu and Koenig, Steven C and Slaughter, Mark S and Giridharan, Guruprasad A},
  journal={ASAIO Journal},
  volume={61},
  number={1},
  pages={21--30},
  year={2015},
  publisher={LWW}
}

@incollection{schmid2020pathophysiological,
  title={Pathophysiological determinants relevant in blood pump control},
  author={Schmid Daners, Marianne and Dual, Seraina Anne},
  booktitle={Mechanical support for heart failure: Current solutions and new technologies},
  pages={253--277},
  year={2020},
  publisher={Springer}
}

@article{bullister2002physiologic,
  title={Physiologic control algorithms for rotary blood pumps using pressure sensor input},
  author={Bullister, Edward and Reich, Sanford and Sluetz, James},
  journal={Artificial organs},
  volume={26},
  number={11},
  pages={931--938},
  year={2002},
  publisher={Wiley Online Library}
}

@article{casas2007minimal,
  title={Minimal sensor count approach to fuzzy logic rotary blood pump flow control},
  author={Casas, Fernando and Ahmed, Nisar and Reeves, Andrew},
  journal={ASAIO Journal},
  volume={53},
  number={2},
  pages={140--146},
  year={2007},
  publisher={LWW}
}

@article{fetanat2019physiological,
  title={A physiological control system for an implantable heart pump that accommodates for interpatient and intrapatient variations},
  author={Fetanat, Masoud and Stevens, Michael and Hayward, Christopher and Lovell, Nigel H},
  journal={IEEE Transactions on Biomedical Engineering},
  volume={67},
  number={4},
  pages={1167--1175},
  year={2019},
  publisher={IEEE}
}

@article{salamonsen2012theoretical,
  title={Theoretical foundations of a Starling-like controller for rotary blood pumps},
  author={Salamonsen, Robert Francis and Lim, Einly and Gaddum, Nicholas and AlOmari, Abdul-Hakeem H and Gregory, Shaun David and Stevens, Michael and Mason, David Glen and Fraser, John F and Timms, Daniel and Karunanithi, Mohan K and others},
  journal={Artificial organs},
  volume={36},
  number={9},
  pages={787--796},
  year={2012},
  publisher={Wiley Online Library}
}

@inproceedings{tanaka2006detection,
  title={Detection and avoiding ventricular suction of ventricular assist devices},
  author={Tanaka, A and Yoshizawa, M and Olegario, P and Ogawa, D and Abe, K and Motomura, T and Igo, S and Nos{\'e}, Y},
  booktitle={2005 IEEE Engineering in Medicine and Biology 27th Annual Conference},
  pages={402--405},
  year={2006},
  organization={IEEE}
}

@article{vollkron2004development,
  title={Development of a suction detection system for axial blood pumps},
  author={Vollkron, Michael and Schima, Heinrich and Huber, Leopold and Benkowski, Robert and Morello, Gino and Wieselthaler, Georg},
  journal={Artificial organs},
  volume={28},
  number={8},
  pages={709--716},
  year={2004},
  publisher={Wiley Online Library}
}

@article{choi2001sensorless,
  title={A sensorless approach to control of a turbodynamic left ventricular assist system},
  author={Choi, Seongjin and Antaki, JE and Boston, R and Thomas, Douglas},
  journal={IEEE Transactions on Control Systems Technology},
  volume={9},
  number={3},
  pages={473--482},
  year={2001},
  publisher={IEEE}
}

@article{parnis1997progress,
  title={Progress in the development of a transcutaneously powered axial flow blood pump ventricular assist system},
  author={Parnis, Steven M and Conger, Jeff L and Fuqua Jr, John M and Jarvik, Robert K and Inman, Rex W and Tamez, Daniel and Macris, Michael P and Moore, Sheila and Jacobs, Gordon and Sweeney, Michael J and others},
  journal={ASAIO Journal},
  volume={43},
  number={5},
  pages={M580},
  year={1997},
  publisher={LWW}
}

@article{waters1999motor,
  title={Motor feedback physiological control for a continuous flow ventricular assist device},
  author={Waters and Allaire and Tao and Bearnson and Wei and Hilton and Baloh and Olsen and Khanwilkar},
  journal={Artificial organs},
  volume={23},
  number={6},
  pages={480--486},
  year={1999},
  publisher={Wiley Online Library}
}

@article{kitamura2000physical,
  title={Physical model-based indirect measurements of blood pressure and flow using a centrifugal pump},
  author={Kitamura, Tadashi and Matsushima, Yuhei and Tokuyama, Toru and Kono, Satoshi and Nishimura, Kazunobu and Komeda, Masahiro and Yanai, Masamichi and Kijima, Toshihiko and Nojiri, Chisato},
  journal={Artificial organs},
  volume={24},
  number={8},
  pages={589--593},
  year={2000},
  publisher={Wiley Online Library}
}

@inproceedings{stevens2011frank,
  title={Frank-starling control of a left ventricular assist device},
  author={Stevens, Michael Charles and Gaddum, Nicholas Richard and Pearcy, Mark and Salamonsen, Robert F and Timms, Daniel Lee and Mason, David Glen and Fraser, John F},
  booktitle={2011 Annual International Conference of the IEEE Engineering in Medicine and Biology Society},
  pages={1335--1338},
  year={2011},
  organization={IEEE}
}

@article{giridharan2003control,
  title={Control strategy for maintaining physiological perfusion with rotary blood pumps},
  author={Giridharan, Guruprasad A and Skliar, Mikhail},
  journal={Artificial organs},
  volume={27},
  number={7},
  pages={639--648},
  year={2003},
  publisher={Wiley Online Library}
}

@article{yazdi2013improvement,
  title={Improvement of Left Ventricular Assist Device (LVAD) in Artificial Heart Using Particle Swarm Optimization},
  author={Yazdi, Majid Neshat and Moghaddam, Reihaneh Kardehi},
  journal={Journal of Artificial Intelligence in Electrical Engineering, Ahar, Iran},
  volume={1},
  number={4},
  pages={9--15},
  year={2013}
}

@article{bakouri2022optimal,
  title={An Optimal H-Infinity Controller for Left Ventricular Assist Devices Based on a Starling-like Controller: A Simulation Study},
  author={Bakouri, Mohsen and Alassaf, Ahmed and Alshareef, Khaled and Abdelsalam, Saleh and Ismail, Husham Farouk and Ganoun, Ali and Alomari, Abdul-Hakeem},
  journal={Mathematics},
  volume={10},
  number={5},
  pages={731},
  year={2022},
  publisher={MDPI}
}

@article{bakouri2014sliding,
  title={A Sliding Mode-Based S tarling-Like Controller for Implantable Rotary Blood Pumps},
  author={Bakouri, Mohsen A and Salamonsen, Robert F and Savkin, Andrey V and AlOmari, Abdul-Hakeem H and Lim, Einly and Lovell, Nigel H},
  journal={Artificial organs},
  volume={38},
  number={7},
  pages={587--593},
  year={2014},
  publisher={Wiley Online Library}
}

@article{bakouri2015nonlinear,
  title={Nonlinear modelling and control of left ventricular assist device},
  author={Bakouri, MA and Savkin, AV and Alomari, AH},
  journal={Electronics Letters},
  volume={51},
  number={8},
  pages={613--615},
  year={2015},
  publisher={Wiley Online Library}
}

@article{son2022model,
  title={Model-Free Adaptive Control of the Failing Heart Managed by Mechanical Supporting Devices},
  author={Son, Jeongeun and Du, Yuncheng},
  journal={IFAC-PapersOnLine},
  volume={55},
  number={7},
  pages={750--755},
  year={2022},
  publisher={Elsevier}
}

@article{chang2011model,
  title={A model-free adaptive control to a blood pump based on heart rate},
  author={Chang, Yu and Gao, Bin and Gu, Kaiyun},
  journal={Asaio Journal},
  volume={57},
  number={4},
  pages={262--267},
  year={2011},
  publisher={LWW}
}

@article{gaddum2014starling,
  title={Starling-like flow control of a left ventricular assist device: in vitro validation},
  author={Gaddum, Nicholas R and Stevens, Michael and Lim, Einly and Fraser, John and Lovell, Nigel and Mason, David and Timms, Daniel and Salamonsen, Robert},
  journal={Artificial organs},
  volume={38},
  number={3},
  pages={E46--E56},
  year={2014},
  publisher={Wiley Online Library}
}

@inproceedings{wang2012feedback,
  title={Feedback control of a rotary left ventricular assist device supporting a failing cardiovascular system},
  author={Wang, Yu and Faragallah, George and Divo, Eduardo and Simaan, Marwan A},
  booktitle={2012 American Control Conference (ACC)},
  pages={1137--1142},
  year={2012},
  organization={IEEE}
}

@article{hornik1989multilayer,
  title={Multilayer feedforward networks are universal approximators},
  author={Hornik, Kurt and Stinchcombe, Maxwell and White, Halbert},
  journal={Neural networks},
  volume={2},
  number={5},
  pages={359--366},
  year={1989},
  publisher={Elsevier}
}

@inproceedings{kim1997control,
  title={Control of left ventricular assist device using artificial neural network},
  author={Kim, Sanghyun and Kim, Hunmo and Ryu, Jungwoo and Chung, Sungtaek},
  booktitle={Proceedings of the 19th Annual International Conference of the IEEE Engineering in Medicine and Biology Society.'Magnificent Milestones and Emerging Opportunities in Medical Engineering'(Cat. No. 97CH36136)},
  volume={3},
  pages={1363--1366},
  year={1997},
  organization={IEEE}
}

@article{ng2017application,
  title={Application of adaptive starling-like controller to total artificial heart using dual rotary blood pumps},
  author={Ng, Boon C and Smith, Peter A and Nestler, Frank and Timms, Daniel and Cohn, William E and Lim, Einly},
  journal={Annals of biomedical engineering},
  volume={45},
  pages={567--579},
  year={2017},
  publisher={Springer}
}

@article{ng2018application,
  title={Application of multiobjective neural predictive control to biventricular assistance using dual rotary blood pumps},
  author={Ng, Boon Chiang and Salamonsen, Robert F and Gregory, Shaun D and Stevens, Michael C and Wu, Yi and Mansouri, Mahdi and Lovell, Nigel H and Lim, Einly},
  journal={Biomedical Signal Processing and Control},
  volume={39},
  pages={81--93},
  year={2018},
  publisher={Elsevier}
}

@article{al2021robustness,
  title={Robustness and performance of deep reinforcement learning},
  author={Al-Nima, Raid Rafi Omar and Han, Tingting and Al-Sumaidaee, Saadoon Awad Mohammed and Chen, Taolue and Woo, Wai Lok},
  journal={Applied Soft Computing},
  volume={105},
  pages={107295},
  year={2021},
  publisher={Elsevier}
}

@article{liu2025multi,
  title={A multi-objective physiological control system for continuous-flow left ventricular assist device with non-invasive physiological feedback},
  author={Liu, Hongtao and Liu, Shuqin},
  journal={Biomedical Signal Processing and Control},
  volume={100},
  pages={107020},
  year={2025},
  publisher={Elsevier}
}

@article{cysyk2019cannula,
  title={Cannula tip with integrated volume sensor for rotary blood pump control: Early-stage development},
  author={Cysyk, Joshua and Newswanger, Ray and Popjes, Eric and Pae, Walter and Jhun, Choon-Sik and Izer, Jenelle and Weiss, William and Rosenberg, Gerson},
  journal={ASAIO journal},
  volume={65},
  number={4},
  pages={318--323},
  year={2019},
  publisher={LWW}
}

@article{ramesh2021considerations,
  title={Considerations to Achieving Pulsatility for Left Ventricular Assist Devices through BLDC Motor by using Closed Loop Control system with PID Controller},
  author={Ramesh, B and Reddy, K Shashidhar and KSV, Phani Kumar and Puppala, Anil Kumar},
  journal={CVR Journal of Science and Technology},
  volume={21},
  number={1},
  pages={69--76},
  year={2021}
}

@article{fetanat2021sensorless,
  title={A sensorless control system for an implantable heart pump using a real-time deep convolutional neural network},
  author={Fetanat, Masoud and Stevens, Michael and Hayward, Christopher and Lovell, Nigel H},
  journal={IEEE Transactions on Biomedical Engineering},
  volume={68},
  number={10},
  pages={3029--3038},
  year={2021},
  publisher={IEEE}
}

@article{fetanat2021fully,
  title={Fully Elman neural network: a novel deep recurrent neural network optimized by an improved Harris Hawks algorithm for classification of pulmonary arterial wedge pressure},
  author={Fetanat, Masoud and Stevens, Michael and Jain, Pankaj and Hayward, Christopher and Meijering, Erik and Lovell, Nigel H},
  journal={IEEE Transactions on Biomedical Engineering},
  volume={69},
  number={5},
  pages={1733--1744},
  year={2021},
  publisher={IEEE}
}

@article{berger2025enhancing,
  title={Enhancing Heart Failure Care: Deep Learning-Based Activity Classification in Left Ventricular Assist Device Patients},
  author={Berger, Laurenz and Haberbusch, Max and Gross, Christoph and Moscato, Francesco},
  journal={ASAIO Journal},
  volume={71},
  number={1},
  pages={52--60},
  year={2025},
  publisher={LWW}
}

@article{just2024artificial,
  title={Artificial intelligence-based analysis of body composition predicts outcome in patients receiving long-term mechanical circulatory support},
  author={Just, Isabell Anna and Schoenrath, Felix and Roehrich, Luise and Heil, Emanuel and Stein, Julia and Auer, Timo Alexander and Fehrenbach, Uli and Potapov, Evgenij and Solowjowa, Natalia and Balzer, Felix and others},
  journal={Journal of Cachexia, Sarcopenia and Muscle},
  volume={15},
  number={1},
  pages={270--280},
  year={2024},
  publisher={Wiley Online Library}
}

@article{bahl2023explainable,
  title={Explainable machine learning analysis of right heart failure after left ventricular assist device implantation},
  author={Bahl, Arjun and Qureshi, Binish and Zhang, Kevin and Bravo, Claudio and Mahr, Claudius and Li, Song},
  journal={ASAIO Journal},
  volume={69},
  number={5},
  pages={417--423},
  year={2023},
  publisher={LWW}
}

@article{raman2004destination,
  title={Destination therapy with ventricular assist devices},
  author={Raman, Jai and Jeevanadam, Valluvan},
  journal={Cardiology},
  volume={101},
  number={1-3},
  pages={104--110},
  year={2004},
  publisher={S. Karger AG}
}

@article{caccamo2011current,
  title={Current state of ventricular assist devices},
  author={Caccamo, Marco and Eckman, Peter and John, Ranjit},
  journal={Current heart failure reports},
  volume={8},
  pages={91--98},
  year={2011},
  publisher={Springer}
}

@article{loor2012pulsatile,
  title={Pulsatile vs. continuous flow in ventricular assist device therapy},
  author={Loor, Gabriel and Gonzalez-Stawinski, Gonzalo},
  journal={Best Practice \& Research Clinical Anaesthesiology},
  volume={26},
  number={2},
  pages={105--115},
  year={2012},
  publisher={Elsevier}
}

@article{jonavsova2021relevance,
  title={On the relevance of boundary conditions and viscosity models in blood flow simulations in patient-specific aorto-coronary bypass models},
  author={Jon{\'a}{\v{s}}ov{\'a}, Alena and Vimmr, Jan},
  journal={International journal for numerical methods in biomedical engineering},
  volume={37},
  number={4},
  pages={e3439},
  year={2021},
  publisher={Wiley Online Library}
}

@article{ortiz2022dynamic,
  title={Dynamic modeling and simulation of the human cardiovascular system with PDA},
  author={Ortiz-Rangel, Estela and Guerrero-Ram{\'\i}rez, Gerardo Vicente and Garc{\'\i}a-Beltr{\'a}n, Carlos Daniel and Guerrero-Lara, Marcela and Adam-Medina, Manuel and Astorga-Zaragoza, Carlos Manuel and Reyes-Reyes, Juan and Posada-G{\'o}mez, Rub{\'e}n},
  journal={Biomedical Signal Processing and Control},
  volume={71},
  pages={103151},
  year={2022},
  publisher={Elsevier}
}

@article{belkacem2023optimizing,
  title={Optimizing Left Ventricular Assist Device Therapy: A Machine Learning Approach for Predicting Cardiac Output},
  author={Belkacem, Marwen and Jemili, Farah and Ellouze, Omar and El Kissi, Asma and Kamel, Ferid},
  year={2023}
}

@book{gregory2024mechanical,
  title={Mechanical circulatory and respiratory support},
  author={Gregory, Shaun D and Stephens, Andrew F and Heinsar, Silver and Arens, Jutta and Fraser, John F},
  year={2024},
  publisher={Elsevier}
}

@article{reesink2007suction,
  title={Suction due to left ventricular assist: implications for device control and management},
  author={Reesink, Koen and Dekker, Andr{\'e} and Van der Nagel, Theo and Beghi, Cesare and Leonardi, Fabio and Botti, Paolo and De Cicco, Giuseppe and Lorusso, Roberto and Van der Veen, Frederik and Maessen, Jos},
  journal={Artificial organs},
  volume={31},
  number={7},
  pages={542--549},
  year={2007},
  publisher={Wiley Online Library}
}

@article{topkara2022machine,
  title={Machine learning-based prediction of myocardial recovery in patients with left ventricular assist device support},
  author={Topkara, Veli K and Elias, Pierre and Jain, Rashmi and Sayer, Gabriel and Burkhoff, Daniel and Uriel, Nir},
  journal={Circulation: Heart Failure},
  volume={15},
  number={1},
  pages={e008711},
  year={2022},
  publisher={Lippincott Williams \& Wilkins Hagerstown, MD}
}

@article{mann2012myocardial,
  title={Myocardial recovery and the failing heart: myth, magic, or molecular target?},
  author={Mann, Douglas L and Barger, Philip M and Burkhoff, Daniel},
  journal={Journal of the American College of Cardiology},
  volume={60},
  number={24},
  pages={2465--2472},
  year={2012},
  publisher={American College of Cardiology Foundation Washington, DC}
}

@article{wilcox2020heart,
  title={Heart failure with recovered left ventricular ejection fraction: JACC scientific expert panel},
  author={Wilcox, Jane E and Fang, James C and Margulies, Kenneth B and Mann, Douglas L},
  journal={Journal of the American College of Cardiology},
  volume={76},
  number={6},
  pages={719--734},
  year={2020},
  publisher={American College of Cardiology Foundation Washington DC}
}

@article{cikes2019machine,
  title={Machine learning-based phenogrouping in heart failure to identify responders to cardiac resynchronization therapy},
  author={Cikes, Maja and Sanchez-Martinez, Sergio and Claggett, Brian and Duchateau, Nicolas and Piella, Gemma and Butakoff, Constantine and Pouleur, Anne Catherine and Knappe, Dorit and Biering-S{\o}rensen, Tor and Kutyifa, Valentina and others},
  journal={European journal of heart failure},
  volume={21},
  number={1},
  pages={74--85},
  year={2019},
  publisher={Wiley Online Library}
}

@article{karantonis2006identification,
  title={Identification and classification of physiologically significant pumping states in an implantable rotary blood pump},
  author={Karantonis, Dean M and Lovell, Nigel H and Ayre, Peter J and Mason, David G and Cloherty, Shaun L},
  journal={Artificial organs},
  volume={30},
  number={9},
  pages={671--679},
  year={2006},
  publisher={Wiley Online Library}
}

@inproceedings{ferreira2006discriminant,
  title={A discriminant-analysis-based suction detection system for rotary blood pumps},
  author={Ferreira, Antonio and Chen, Shaohui and Simaan, Marwan A and Boston, J Robert and Antaki, James F},
  booktitle={2006 International Conference of the IEEE Engineering in Medicine and Biology Society},
  pages={5382--5385},
  year={2006},
  organization={IEEE}
}

@article{karantonis2008noninvasive,
  title={Noninvasive detection of suction in an implantable rotary blood pump using neural networks},
  author={Karantonis, Dean M and Cloherty, Shaun L and Lovell, Nigel H and Mason, David G and Salamonsen, Robert F and Ayre, Peter J},
  journal={International Journal of Computational Intelligence and Applications},
  volume={7},
  number={03},
  pages={237--247},
  year={2008},
  publisher={World Scientific}
}

@article{wang2013suction,
  title={A suction detection system for rotary blood pumps based on the Lagrangian support vector machine algorithm},
  author={Wang, Yu and Simaan, Marwan A},
  journal={IEEE Journal of Biomedical and Health Informatics},
  volume={17},
  number={3},
  pages={654--663},
  year={2013},
  publisher={IEEE}
}

@article{aly2011pid,
  title={PID parameters optimization using genetic algorithm technique for electrohydraulic servo control system},
  author={Aly, Ayman A and others},
  journal={Intelligent Control and Automation},
  volume={2},
  number={02},
  pages={69},
  year={2011},
  publisher={Scientific Research Publishing}
}

@article{ayala2012tuning,
  title={Tuning of PID controller based on a multiobjective genetic algorithm applied to a robotic manipulator},
  author={Ayala, Helon Vicente Hultmann and dos Santos Coelho, Leandro},
  journal={Expert Systems with Applications},
  volume={39},
  number={10},
  pages={8968--8974},
  year={2012},
  publisher={Elsevier}
}

@article{ketelhut2017iterative,
  title={Iterative learning control of a left ventricular assist device},
  author={Ketelhut, Maike and Schr{\"o}del, Frank and Stemmler, Sebastian and Roseveare, Jesse and Hein, Marc and Gesenhues, Jonas and Albin, Thivarian and Abel, Dirk},
  journal={IFAC-PapersOnLine},
  volume={50},
  number={1},
  pages={6684--6690},
  year={2017},
  publisher={Elsevier}
}

@article{ruschen2017minimizing,
  title={Minimizing left ventricular stroke work with iterative learning flow profile control of rotary blood pumps},
  author={R{\"u}schen, Daniel and Prochazka, Frederik and Amacher, Raffael and Bergmann, Lukas and Leonhardt, Steffen and Walter, Marian},
  journal={Biomedical Signal Processing and Control},
  volume={31},
  pages={444--451},
  year={2017},
  publisher={Elsevier}
}

@article{ketelhut2018iterative,
  title={Iterative learning control of a left ventricular assist device: Nonlinear model integration},
  author={Ketelhut, M and Stemmler, S and Hein, M and K{\"o}rner, D and Abel, D},
  journal={IFAC-PapersOnLine},
  volume={51},
  number={27},
  pages={152--157},
  year={2018},
  publisher={Elsevier}
}

@article{ketelhut2019iterative,
  title={Iterative learning control of ventricular assist devices with variable cycle durations},
  author={Ketelhut, Maike and Stemmler, Sebastian and Gesenhues, Jonas and Hein, Marc and Abel, Dirk},
  journal={Control Engineering Practice},
  volume={83},
  pages={33--44},
  year={2019},
  publisher={Elsevier}
}

@article{magkoutas2022physiologic,
  title={Physiologic data-driven iterative learning control for left ventricular assist devices},
  author={Magkoutas, Konstantinos and Arm, Philip and Meboldt, Mirko and Schmid Daners, Marianne},
  journal={Frontiers in Cardiovascular Medicine},
  volume={9},
  pages={922387},
  year={2022},
  publisher={Frontiers}
}

@article{magkoutas2023genetic,
  title={Genetic algorithm-based optimization framework for control parameters of ventricular assist devices},
  author={Magkoutas, Konstantinos and Rossato, Leonardo Nunes and Heim, Marco and Daners, Marianne Schmid},
  journal={Biomedical Signal Processing and Control},
  volume={85},
  pages={104788},
  year={2023},
  publisher={Elsevier}
}

@article{howell2021using,
  title={Using machine-learning for prediction of the response to cardiac resynchronization therapy: the SMART-AV study},
  author={Howell, Stacey J and Stivland, Tim and Stein, Kenneth and Ellenbogen, Kenneth A and Tereshchenko, Larisa G},
  journal={Clinical Electrophysiology},
  volume={7},
  number={12},
  pages={1505--1515},
  year={2021},
  publisher={American College of Cardiology Foundation Washington DC}
}

@article{de2023machine,
  title={A machine learning method integrating ECG and gated SPECT for cardiac resynchronization therapy decision support},
  author={de A. Fernandes, Fernando and Larsen, Kristoffer and He, Zhuo and Nascimento, Erivelton and Peix, Amalia and Sha, Qiuying and Paez, Diana and Garcia, Ernest V and Zhou, Weihua and Mesquita, Claudio T},
  journal={European journal of nuclear medicine and molecular imaging},
  volume={50},
  number={10},
  pages={3022--3033},
  year={2023},
  publisher={Springer}
}

@article{park2024machine,
  title={A Machine Learning-derived Risk Score Improves Prediction of Outcomes After LVAD Implantation: An Analysis of the INTERMACS Database},
  author={Park, Jin Joo and John, Sonya and Campagnari, Claudio and Yagil, Avi and Greenberg, Barry and Adler, Eric},
  journal={Journal of Cardiac Failure},
  year={2024},
  publisher={Elsevier}
}

\end{document}